\documentclass[journal]{IEEEtran}

\usepackage[T1]{fontenc}
\usepackage{amsmath,amssymb,amsfonts}
\usepackage{graphicx}
\usepackage{textcomp}
\usepackage[table]{xcolor}
\usepackage{booktabs}
\usepackage{array}
\usepackage{multirow}
\usepackage{longtable}
\usepackage{tabularx}
\usepackage{adjustbox}
\usepackage{rotating}
\usepackage{pdflscape}
\usepackage{threeparttable}
\usepackage{url}
\usepackage{listings}
\usepackage{tikz}
\usetikzlibrary{shapes.geometric,arrows.meta,positioning,fit,backgrounds,calc}
\usepackage{placeins}
\usepackage{flafter}
\usepackage{dblfloatfix}
\usepackage{afterpage}
\usepackage[colorlinks=true,linkcolor=black,citecolor=black,urlcolor=blue]{hyperref}
\newcommand{\studyref}[2]{\textbf{#1}\,\cite{#2}}
\begin{document}

\title{From Data to Deployment: A Scoping Review of Empirical Tools, Techniques, and Operational Readiness in Data-Driven 5G/6G Network Optimization}

\author{Muhammad~Kabeer, Rosdiadee~Nordin, and Sian~Lun~Lau%
\thanks{Muhammad Kabeer and Sian Lun Lau are with the Faculty of Engineering and Technology, Sunway University, Bandar Sunway 47500, Selangor, Malaysia. Muhammad Kabeer is also with the Department of Computer Science, Federal University Dutsinma, Katsina, Nigeria. Sian Lun Lau is also with the School of Computer Science and Engineering, Vellore Institute of Technology, Chennai, India.}%
\thanks{Rosdiadee Nordin is with the Future Cities Research Institute, Faculty of Engineering and Technology, Sunway University, Bandar Sunway 47500, Selangor, Malaysia, and the Future Cities Research Institute, Lancaster University, Lancaster LA1 4YW, United Kingdom (corresponding author: rosdiadeen@sunway.edu.my).}}

\maketitle

\begin{abstract}
Artificial intelligence (AI) is increasingly applied to 5G/6G network optimization, yet differences in empirical evidence, evaluation methods, and computational requirements complicate comparisons of reported performance. This scoping review synthesizes 72 peer-reviewed journal studies published between 2022 and 2026, selected through a PRISMA-ScR-informed process from 893 records. The review examines empirical tools, AI/ML techniques, optimization domains, and operational readiness. Evidence classification identifies 37 studies (51.4\%) using real-world or physical-testbed evidence only, 25 (34.7\%) combining measured and simulated evidence, and 10 (13.9\%) relying exclusively on synthetic or simulated data. Offline analysis was the most common evaluation setting, accounting for 45 studies (62.5\%). Three principal findings emerge. First, empirical measurements improve contextual relevance, while simulation and digital twins support scalable experimentation but introduce validation challenges. Second, algorithmic effectiveness depends on the optimization objective, data characteristics, and computational constraints rather than model complexity alone. Third, predictive accuracy is more consistently evaluated than the network-level effects of optimization decisions. The review therefore introduces the six-module Evidence Fidelity and Operational Reporting (EFOR) checklist to improve evaluation consistency, reporting transparency, and operational relevance.
\end{abstract}


\section{Introduction}

\subsection{Motivation}

As cellular networks evolve from fifth generation (5G) toward sixth generation (6G), operators and researchers are exploring how artificial intelligence (AI) can complement established threshold-based controls. Applications that depend on stable throughput, low latency, and reliable mobility create a clear need to anticipate congestion, coverage degradation, and handover problems. Many methods report encouraging predictive results, but far fewer show how those results translate into sustained network operation.

Existing reviews address different parts of data-driven 5G and 6G optimization. Islam~et al.~\cite{ISLAM2026390} provide the closest methodological comparison by linking 6G key performance indicators (KPIs) with AI-based performance evaluation, optimization methods, and evaluation tools. Their evidence is predominantly simulated, with limited alignment among benchmarks. Jiang~\cite{JIANG2022117163} develops taxonomies of datasets, models, workflows, and benchmarks for cellular traffic prediction and identifies a persistent gap between research and deployment. Tang~et al.~\cite{9403380} review machine learning (ML) for network access, routing, congestion control, and adaptive streaming, with attention to computation, transferability, benchmarking, and access to open resources. Mustafovski~et al.~\cite{mustafovski2026aidriven} provide a taxonomy spanning 5G and 6G, normalized KPIs, and evidence of deployment maturity. Singh~et al.~\cite{SINGH2025111663} examine 6G KPIs, enabling and integrated technologies, implementation challenges, and links between academia and industry. Lykakis~et al.~\cite{lykakis2025data} focus on data traffic prediction for 5G and beyond, including prediction challenges, dataset characteristics, and methods that protect privacy. Nurcahyanto~et al.~\cite{nurcahyanto2025systematic} examine ML for telecommunications reliability, particularly fault prediction, maintenance, and explainability. Zhang and Zhu~\cite{zhang2020towards} give a broader account of AI-enabled 6G networking across radio systems, traffic control, security, and orchestration. Roessler~et al.~\cite{11404155} discuss AI and ML solutions for 6G networks and their energy implications through the 6G Access, Network of Networks, Automation and Simplification (6G-ANNA) initiative.

Taken together, these reviews explain important parts of AI-enabled network optimization, but they seldom connect the origin of the evidence, the algorithmic role, the optimization domain, the reported numerical result, and the conditions of deployment in one analysis. A metric improvement is most informative when the reader can also see how the data were collected, where the method was evaluated, and what kind of network decision followed. This review makes those connections across drive-test tools, operator KPI feeds, simulation and emulation platforms, digital twins, and physical testbeds.

The practical value of a model depends on more than its predictive accuracy. The observations need a clear provenance, the evaluation should reflect the intended operating conditions, and the resulting estimate must lead to a feasible decision whose effects can be monitored. Measured traffic, for example, may support an offline forecasting result without showing that a resource-allocation action will be timely or safe. A programmable testbed can demonstrate execution, yet still leave transfer across operators, vendors, and changing traffic conditions unresolved. These differences shape the research questions below and the interpretation of the evidence throughout the review.

\subsection{Objective}

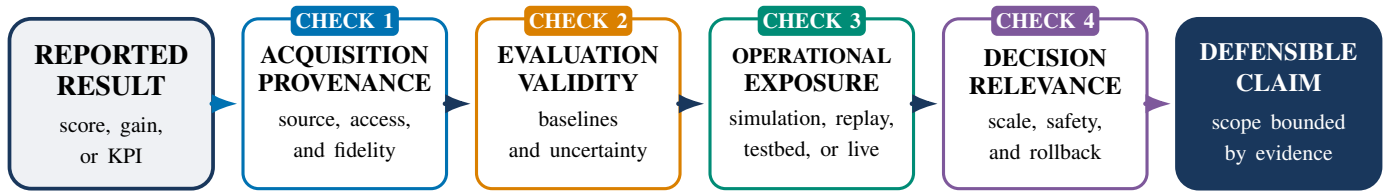
\begin{figure*}[!tbp]
\centering
\resizebox{\textwidth}{!}{%
\begin{tikzpicture}[font=\small,>=Latex]
\definecolor{auditnavy}{RGB}{25,55,92}
\definecolor{auditblue}{RGB}{0,114,178}
\definecolor{auditorange}{RGB}{218,128,0}
\definecolor{auditteal}{RGB}{0,140,118}
\definecolor{auditpurple}{RGB}{126,88,155}

\path[use as bounding box] (-7.75,-1.05) rectangle (7.75,1.48);

\tikzset{
  endpoint/.style={rounded corners=7pt,line width=1.2pt,minimum height=1.92cm,
    text width=1.95cm,align=center,inner sep=5pt},
  auditgate/.style={rounded corners=5pt,line width=1pt,fill=white,
    text width=2.02cm,minimum height=1.92cm,align=center,inner sep=4pt},
  checktag/.style={rounded corners=2pt,text=white,font=\bfseries\scriptsize,
    minimum width=1.02cm,minimum height=.34cm,inner sep=1.5pt}
}

\node[endpoint,draw=auditnavy,fill=auditnavy!7] (result) at (-6.60,.26)
  {\textbf{REPORTED\\RESULT}\\[4pt]\scriptsize score, gain,\\or KPI};

\node[auditgate,draw=auditblue] (g1) at (-3.96,.26)
  {{\bfseries\footnotesize ACQUISITION}\\[-1pt]{\bfseries\footnotesize PROVENANCE}\\[3pt]\scriptsize source, access,\\and fidelity};
\node[checktag,fill=auditblue] at ([yshift=-.04cm]g1.north) {CHECK 1};

\node[auditgate,draw=auditorange] (g2) at (-1.32,.26)
  {{\bfseries\footnotesize EVALUATION}\\[-1pt]{\bfseries\footnotesize VALIDITY}\\[3pt]\scriptsize baselines\\and uncertainty};
\node[checktag,fill=auditorange] at ([yshift=-.04cm]g2.north) {CHECK 2};

\node[auditgate,draw=auditteal] (g3) at (1.32,.26)
  {{\bfseries\scriptsize OPERATIONAL}\\[-1pt]{\bfseries\footnotesize EXPOSURE}\\[3pt]\scriptsize simulation, replay,\\testbed, or live};
\node[checktag,fill=auditteal] at ([yshift=-.04cm]g3.north) {CHECK 3};

\node[auditgate,draw=auditpurple] (g4) at (3.96,.26)
  {{\bfseries\footnotesize DECISION}\\[-1pt]{\bfseries\footnotesize RELEVANCE}\\[3pt]\scriptsize scale, safety,\\and rollback};
\node[checktag,fill=auditpurple] at ([yshift=-.04cm]g4.north) {CHECK 4};

\node[endpoint,draw=auditnavy,fill=auditnavy,text=white] (claim) at (6.60,.26)
  {{\bfseries\footnotesize DEFENSIBLE\\CLAIM}\\[4pt]\scriptsize scope bounded\\by evidence};

\foreach \a/\b/\c in {result/g1/auditblue,g1/g2/auditnavy,g2/g3/auditnavy,g3/g4/auditnavy,g4/claim/auditpurple}{
  \draw[-{Latex[length=3.0mm,width=2.2mm]},line width=1.5pt,line cap=round,shorten <=1pt,shorten >=1pt,\c]
    (\a.east) -- (\b.west);
}

\end{tikzpicture}%
}
\caption{Evidence checks for deployment claims.}
\label{fig:operator_workflow}
\end{figure*}

This scoping review examines the empirical tools, algorithmic techniques, and optimization-domain interventions reported in 72 peer-reviewed journal articles published from January 2022 to June 2026. Instead of ranking results obtained from incompatible datasets and protocols, it asks how field measurements, crowdsourced records, operator feeds, simulation, and testbeds affect the conclusions that can be drawn from different algorithms. Figure~\ref{fig:operator_workflow} organizes this assessment around acquisition provenance, evaluation validity, operational exposure, and decision relevance. The analysis shows a recurring trade-off between contextual fidelity and scalable data collection, identifies task-dependent strengths and constraints across algorithm families, and traces the evidence needed to connect validated observations with network action and subsequent monitoring.

\subsection{Scope and contributions}
\label{sec:intro_rqs}

To address the operational gaps identified in the existing literature, this scoping review considers three research questions.

\begin{itemize}
\item \textbf{RQ1. Empirical Tools.} How do current empirical data acquisition tools and testbeds impact the fidelity and scalability of data collection for cellular network optimization?
\item \textbf{RQ2. Algorithmic Techniques.} How do different algorithmic families contribute to proactive network management, and what operational limitations do they introduce in 5G/6G environments?
\item \textbf{RQ3. Optimization Domains.} How can data-driven interventions be applied across optimization domains to build resilient, self-optimizing network infrastructure?
\end{itemize}

The review makes three scientific contributions.

\begin{enumerate}
\item An empirical-tool and testbed taxonomy that establishes the trade-off between contextual fidelity and scalable, accessible, and reproducible data collection.
\item A task-aware comparison showing that algorithmic families provide complementary management capabilities whose operational value depends on validation, computational cost, coordination overhead, and deployment context.
\item A cross-domain synthesis that defines the observation-to-action requirements for resilient, self-optimizing networks and translates the remaining evidence gaps into the proposed six-module EFOR checklist.
\end{enumerate}

Table~\ref{tab:related_reviews} compares the review with nine closely related surveys across five substantive evidence-synthesis dimensions. A checkmark requires a structured synthesis that directly addresses the stated dimension; descriptive discussion or isolated examples alone do not qualify.

\begin{table*}[!tbp]
\centering
\caption{Comparison with Related Reviews Across Evidence-Synthesis Dimensions}
\label{tab:related_reviews}
\renewcommand{\arraystretch}{1.3}
\begingroup
\setlength{\tabcolsep}{3pt}
\scriptsize
\resizebox{\textwidth}{!}{%
\begin{tabular}{@{}>{\raggedright\arraybackslash}p{3.2cm}*{5}{>{\centering\arraybackslash}p{2.15cm}}@{}}
\toprule
\multirow{2}{*}{\textbf{Study}} & \textbf{Acquisition Provenance\newline and Fidelity Trade-offs} & \textbf{Algorithms Linked to\newline Operational Constraints} & \textbf{Cross-domain\newline Intervention Synthesis} & \textbf{Corpus-level\newline Quantitative\newline Synthesis} & \textbf{Operational-evidence\newline Stratification} \\
\midrule
Islam~et al. (2026)~\cite{ISLAM2026390} & & \checkmark & \checkmark & \checkmark & \\
\addlinespace
Jiang (2022)~\cite{JIANG2022117163} & & \checkmark & & \checkmark & \\
\addlinespace
Tang~et al. (2021)~\cite{9403380} & & \checkmark & \checkmark & \checkmark & \\
\addlinespace
Mustafovski~et al. (2026)~\cite{mustafovski2026aidriven} & & \checkmark & \checkmark & \checkmark & \checkmark \\
\addlinespace
Singh~et al. (2025)~\cite{SINGH2025111663} & & \checkmark & \checkmark & & \\
\addlinespace
Lykakis~et al. (2025)~\cite{lykakis2025data} & & \checkmark & & \checkmark & \\
\addlinespace
Nurcahyanto~et al. (2025)~\cite{nurcahyanto2025systematic} & & \checkmark & & \checkmark & \\
\addlinespace
Zhang and Zhu (2020)~\cite{zhang2020towards} & & \checkmark & \checkmark & & \\
\addlinespace
Roessler~et al. (2026)~\cite{11404155} & & \checkmark & \checkmark & & \\
\addlinespace
\textbf{This scoping review (2026)} & \checkmark & \checkmark & \checkmark & \checkmark & \checkmark \\
\bottomrule
\end{tabular}
}
\endgroup
\end{table*}

Section~\ref{sec:methodology} describes the search strategy, eligibility criteria, and screening outcomes. Section~\ref{sec:taxonomy} presents the taxonomy of empirical tools and data acquisition methods. Section~\ref{sec:algorithms} compares the main algorithmic techniques, and Section~\ref{sec:domains} maps them to optimization domains and interventions. Section~\ref{sec:synthesis} integrates the evidence across themes and exposure levels. Section~\ref{sec:gaps} develops the resulting challenges, reporting needs, and research directions, before Section~\ref{sec:conclusion} concludes the paper.

\section{Scoping review methodology}
\label{sec:methodology}

This scoping review maps the range, characteristics, and operational implications of a heterogeneous evidence base. Study identification and reporting were informed by PRISMA-ScR~\cite{Tricco2018PRISMAScR}, with consistent eligibility and appraisal criteria across the four databases.

The search boundary was English-language, peer-reviewed journal research articles published from 2022 through 2026, retrieved from IEEE Xplore, the ACM Digital Library, Elsevier ScienceDirect, or Scopus, and reporting quantified data-driven optimization of terrestrial cellular networks relevant to 5G/6G management.

\subsection{Research questions driving the search}

We derived the search strategy and eligibility criteria directly from the three research questions introduced in Section~\ref{sec:intro_rqs}. The search therefore covered empirical acquisition tools, testbeds, source provenance, collection fidelity and scale, algorithmic families and their management roles, operational limitations, optimization domains, and evidence relevant to resilient, self-optimizing infrastructure, including edge-computing and digital-twin integration.

\subsection{Search strategy}

We searched IEEE Xplore, the ACM Digital Library, Elsevier ScienceDirect, and Scopus using database-specific Boolean expressions covering network generation, empirical data acquisition, learning or prediction, and cellular-network optimization. The database-specific search expressions, including the IEEE search fields and Scopus TITLE-ABS-KEY qualifier, are presented alongside the screening counts in Table~\ref{tab:screening}. The literature search was conducted on 28 June 2026. Table~\ref{tab:screening} reports the search and screening totals for the 72 retained articles.

Because IEEE Xplore is a major source of cellular-network research, additional cellular, empirical, and AI synonyms were included to avoid missing relevant studies indexed under alternative terminology. Table~\ref{tab:screening} reports the initial results, records remaining after the language, year, and article-type restrictions, records advanced after title and abstract screening, and eligible full-text articles retained.

\begin{table*}[!tbp]
\centering
\caption{Database search queries and screening outcomes. Search expressions are reproduced verbatim; line breaks are typographic.}
\label{tab:screening}
\footnotesize
\setlength{\tabcolsep}{4pt}
\renewcommand{\arraystretch}{1.18}
\begin{tabularx}{\textwidth}{@{}>{\raggedright\arraybackslash}p{2.15cm}>{\raggedright\arraybackslash}X*{4}{>{\centering\arraybackslash}p{1.45cm}}@{}}
\toprule
\textbf{Database} & \textbf{Search Query} & \shortstack{\textbf{Initial}\\\textbf{Search}} & \shortstack{\textbf{After}\\\textbf{Filter}} & \shortstack{\textbf{Title and}\\\textbf{Abstract}} & \shortstack{\textbf{Full}\\\textbf{Text}} \\
\midrule
Elsevier ScienceDirect & {\raggedright\ttfamily\scriptsize\sloppy (\textquotedbl{}5G\textquotedbl{} OR \textquotedbl{}6G\textquotedbl{}) AND (\textquotedbl{}drive test\textquotedbl{} OR \textquotedbl{}real world data\textquotedbl{}) AND (\textquotedbl{}machine learning\textquotedbl{} OR \textquotedbl{}prediction\textquotedbl{}) AND (\textquotedbl{}cellular network optimization\textquotedbl{} OR \textquotedbl{}cellular networks\textquotedbl{})} & 188 & 82 & 23 & 13 \\
\addlinespace[3pt]
ACM Digital Library & {\raggedright\ttfamily\scriptsize\sloppy (\textquotedbl{}5G\textquotedbl{} OR \textquotedbl{}6G\textquotedbl{}) AND (\textquotedbl{}drive test\textquotedbl{} OR \textquotedbl{}real world data\textquotedbl{}) AND (\textquotedbl{}machine learning\textquotedbl{} OR \textquotedbl{}prediction\textquotedbl{}) AND (\textquotedbl{}cellular network optimization\textquotedbl{} OR \textquotedbl{}cellular networks\textquotedbl{})} & 171 & 17 & 5 & 4 \\
\addlinespace[3pt]
IEEE Xplore & {\raggedright\ttfamily\scriptsize\sloppy (Document Title,Abstract,Keywords: \textquotedbl{}5G\textquotedbl{} OR \textquotedbl{}6G\textquotedbl{} OR \textquotedbl{}cellular\textquotedbl{}) AND (Document Title,Abstract,Keywords: \textquotedbl{}drive test\textquotedbl{} OR \textquotedbl{}real world\textquotedbl{} OR \textquotedbl{}empirical\textquotedbl{} OR \textquotedbl{}measurement\textquotedbl{}) AND (Document Title,Abstract,Keywords: \textquotedbl{}machine learning\textquotedbl{} OR \textquotedbl{}predictive\textquotedbl{} OR \textquotedbl{}AI\textquotedbl{}) AND (Document Title,Abstract,Keywords: \textquotedbl{}optimization\textquotedbl{} OR \textquotedbl{}management\textquotedbl{})} & 441 & 113 & 42 & 29 \\
\addlinespace[3pt]
Scopus & {\raggedright\ttfamily\scriptsize\sloppy TITLE-ABS-KEY ( ( \textquotedbl{}5G\textquotedbl{} OR \textquotedbl{}6G\textquotedbl{} ) AND ( \textquotedbl{}drive test\textquotedbl{} OR \textquotedbl{}drive-test\textquotedbl{} OR \textquotedbl{}real-world data\textquotedbl{} OR \textquotedbl{}real-world\textquotedbl{} ) AND ( \textquotedbl{}machine learning\textquotedbl{} OR \textquotedbl{}prediction\textquotedbl{} ) AND ( \textquotedbl{}cellular network\textquotedbl{} OR \textquotedbl{}network planning\textquotedbl{} ) )} & 93 & 39 & 33 & 26 \\
\addlinespace[3pt]
\midrule
\textbf{Total} & --- & \textbf{893} & \textbf{251} & \textbf{103} & \textbf{72} \\
\bottomrule
\end{tabularx}
\end{table*}

\begin{figure*}[!tbp]
\centering
\begin{minipage}[c]{0.58\textwidth}
\centering
\includegraphics[width=\linewidth]{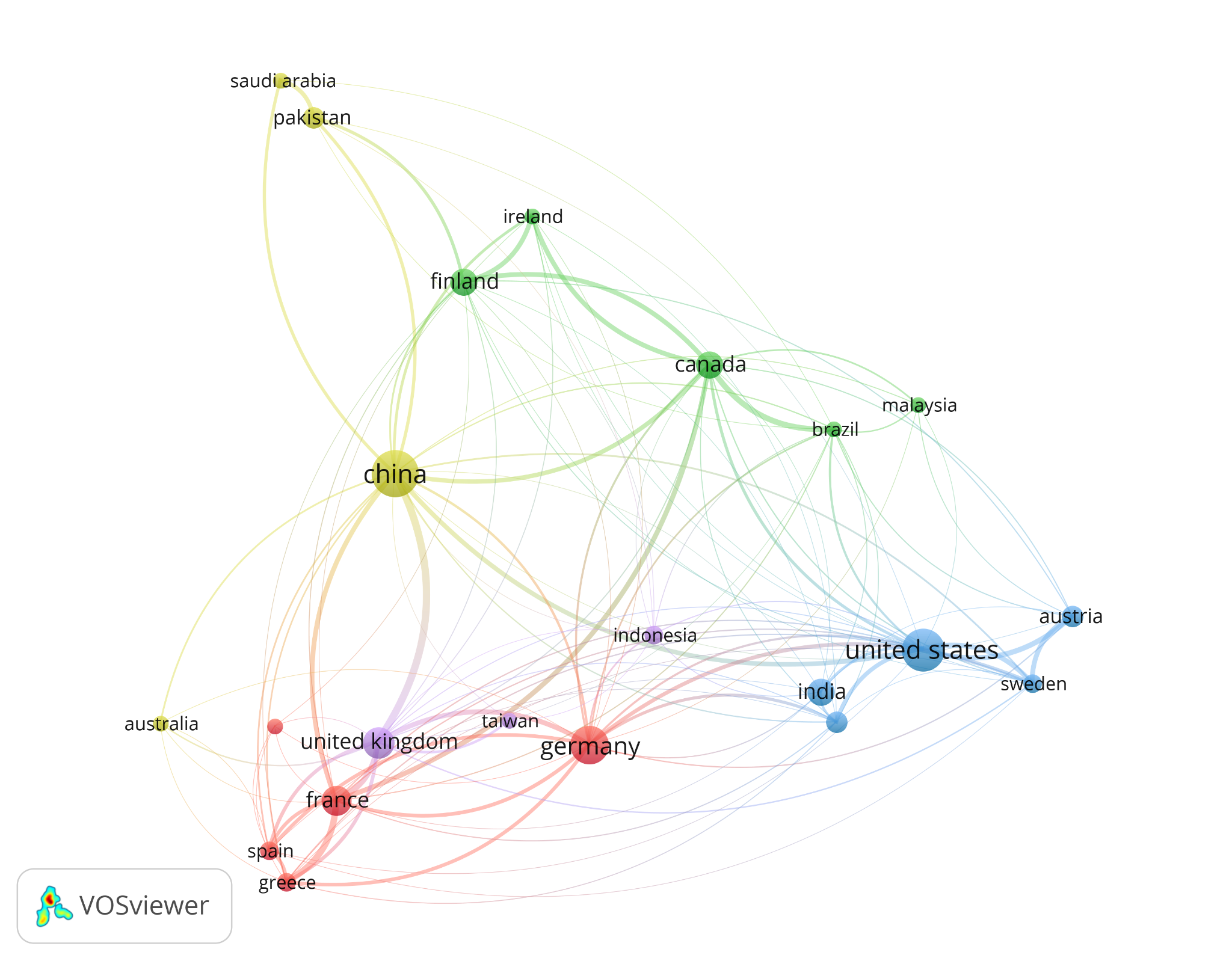}
\textbf{(a)} Country co-authorship
\end{minipage}\hfill
\begin{minipage}[c]{0.38\textwidth}
\centering
\includegraphics[width=\linewidth]{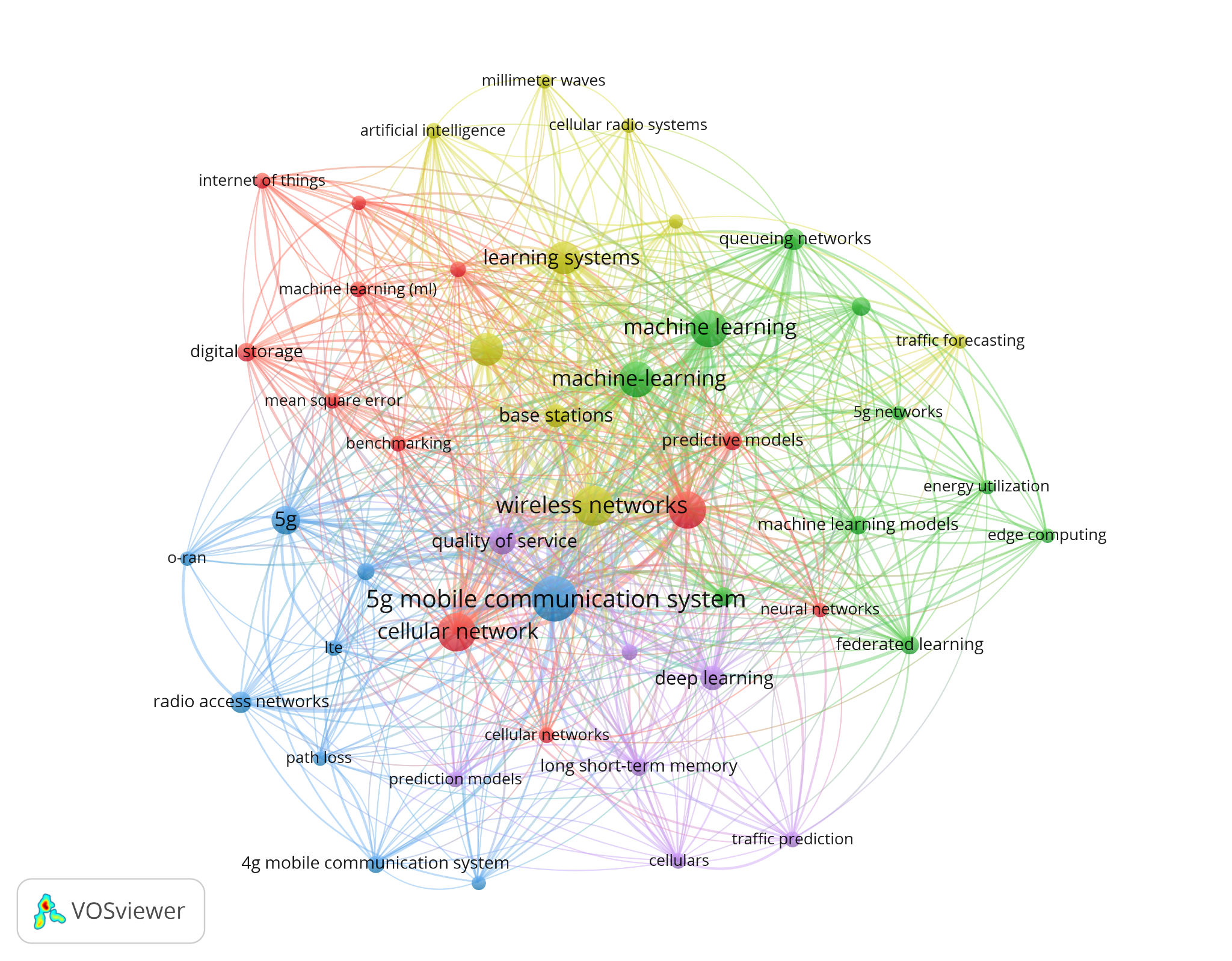}
\textbf{(b)} Keyword co-occurrence
\end{minipage}
\caption{VOSviewer maps of the Scopus search results.}
\label{fig:vos_combined}
\end{figure*}

VOSviewer was used to characterize the Scopus search results (Figure~\ref{fig:vos_combined}). Scopus was used as the representative source for this bibliometric view because its broad indexed coverage substantially overlaps with publications retrieved through IEEE Xplore, the ACM Digital Library, and Elsevier ScienceDirect. The country network shows internationally distributed collaboration, with prominent nodes for the United States, China, Germany, the United Kingdom, France, and Canada. The keyword network connects cellular infrastructure with machine learning, traffic prediction, edge computing, federated learning, and Open Radio Access Network (O-RAN) research. Together, the maps indicate the geographic concentration and thematic direction of the wider research landscape, providing context for the trends and research priorities examined in the retained studies.

\subsection{Inclusion and exclusion criteria}

Table~\ref{tab:inclusion} summarizes the full-text eligibility criteria. Included studies reported an identifiable evidence source and a quantified method linked to terrestrial cellular-network management or optimization. Prediction studies qualified only when their outputs were explicitly connected to an optimization decision. Non-terrestrial studies were excluded unless they contributed directly to terrestrial optimization; UAVs used only as measurement platforms remained eligible.

\begin{table}[!tbp]
\centering
\caption{Inclusion and Exclusion Criteria}
\label{tab:inclusion}
\small
\renewcommand{\arraystretch}{1.25}
\begin{tabularx}{\columnwidth}{@{}>{\raggedright\arraybackslash}p{1.8cm}>{\raggedright\arraybackslash}X@{}}
\toprule
\textbf{Criterion} & \textbf{Description} \\
\midrule
\multicolumn{2}{@{}l}{\textit{Inclusion}} \\
I1 & Addresses 5G, 6G, a transitional generation, or a terrestrial cellular-management problem directly relevant to 5G/6G optimization. \\
I2 & Reports field/operator data, a physical testbed, or a quantified synthetic, simulated, or emulated evidence source. \\
I3 & Applies a quantified data-driven, statistical, machine-learning, deep-learning, reinforcement-learning, or model-based method to a network-management decision or an explicitly stated optimization input, with reported performance metrics. \\
I4 & Published as a peer-reviewed journal article from 2022 to 2026. \\
I5 & Published in English. \\
\midrule
\multicolumn{2}{@{}l}{\textit{Exclusion}} \\
E1 & Purely conceptual studies with no quantified empirical, testbed, simulation, or emulation evaluation. \\
E2 & Surveys, tutorials, position papers, magazines, conference papers/proceedings, or other non-primary/non-journal publication types. \\
E3 & Domain mismatch, including satellite/UAV-based NTN studies without a direct terrestrial cellular-optimization contribution. \\
E4 & Duplicate publications or extended abstracts of an already-included study. \\
\bottomrule
\end{tabularx}
\end{table}

For I1, an explicit 5G or 6G label was sufficient but not required. The two LTE-only studies and the 24 generation-agnostic studies were retained only when they examined a terrestrial cellular-management function directly relevant to 5G/6G optimization and satisfied I2 through I5; the generation-specific sensitivity analysis reports the effect of excluding these studies.

For I3, an ``explicitly stated optimization input'' required the paper to identify the downstream planning, configuration, association, allocation, or control decision informed by the prediction. A generic claim that improved prediction could aid optimization was insufficient. Signal and path-loss studies were therefore retained only when their stated use included coverage planning, beam or cell selection, trajectory design, or another specified network decision.

\subsection{Screening and eligibility assessment}

One reviewer screened the titles, abstracts, and 103 full-text reports using ASReview LAB to support record prioritization and organization. The co-authors closely supervised the application of the eligibility criteria and reviewed uncertain decisions to reduce classification bias. Independent duplicate screening, adjudication counts, and inter-rater agreement statistics were not recorded. Co-author supervision is therefore not equivalent to independently replicated screening, and the categorization results should be interpreted with that limitation. Ambiguous title or abstract records were retained for full-text assessment. The final venue check excluded three IEEE magazine articles, leaving 72 studies: 13 from Elsevier ScienceDirect, 4 from the ACM Digital Library, 29 from IEEE Xplore, and 26 from Scopus. Figure~\ref{fig:prisma} summarizes the selection process.

\begin{figure}[!tbp]
\centering
\begin{tikzpicture}[
    font=\scriptsize,
    box/.style={draw, rectangle, rounded corners=1pt, align=center, minimum width=4.0cm, minimum height=0.9cm, fill=blue!6, inner sep=3pt},
    excl/.style={draw, rectangle, align=left, text width=3.0cm, minimum width=3.0cm, fill=red!6, inner sep=2pt, font=\fontsize{5.8}{6.4}\selectfont},
    arrow/.style={-{Triangle[length=2mm,width=2mm]}, thick},
    node distance=7mm
]

\node[box] (id1) {\textbf{Identification}\\IEEE Xplore, ACM Digital Library,\\Elsevier ScienceDirect, Scopus\\($n = 893$)};

\node[box, below=of id1] (filt) {Records after filters\\($n = 251$)};
\draw[arrow] (id1) -- (filt);

\node[box, below=of filt] (tasc) {Records screened by title and abstract\\($n = 251$)};
\draw[arrow] (filt) -- (tasc);

\node[excl, right=3mm of tasc] (exc0) {Records removed during title and abstract screening ($n=148$):\\Duplicates identified ($n=7$)\\Excluded ($n=141$)};
\draw[arrow] (tasc) -- (exc0);

\node[box, below=of tasc] (elig) {Full-text articles assessed for eligibility\\($n = 103$)};
\draw[arrow] (tasc) -- (elig);

\node[excl, right=3mm of elig] (exc1) {Full-text articles excluded under E1--E4 ($n=31$)};
\draw[arrow] (elig) -- (exc1);

\node[box, below=of elig, fill=green!12] (final) {\textbf{Studies included in synthesis}\\Elsevier 13 + ACM 4 + IEEE 29 + Scopus 26\\($n = 72$)};
\draw[arrow] (elig) -- (final);

\end{tikzpicture}
\caption{PRISMA-ScR-Informed Study-Selection Flow}
\label{fig:prisma}
\end{figure}

The primary reviewer assessed and classified the included publications, with the co-authors supervising uncertain decisions. The comparative appraisal considered bibliographic context, evidence source and accessibility, observation scale, evaluation setting, algorithmic role, optimization task, comparator, metric, computational reporting, deployment conditions, limitations, and inconsistencies. Evidence was classified as measured or testbed, generated, or hybrid. Fidelity was graded separately as generated evidence, legacy or proxy evidence, contemporary real-world evidence, or physical testbed/over-the-air evidence. Each study also received one primary theme and one highest exposure level: offline analysis, replay or emulation, simulation or synthetic evaluation, physical implementation or measurement, or sustained live control.

The exposure decision followed a separate acquisition--action rule. Real measurements were recorded independently of whether the evaluated model ran during collection. Offline analysis covered models applied after data collection, including TEMS, Nemo, crowdsourced, and operator-log studies. Physical implementation or measurement required the evaluated procedure or model to operate during a physical experiment; it did not by itself imply a network-management intervention. A separate indicator recorded whether the evaluated procedure executed an action on real equipment. Operator affiliation or access to production logs did not raise exposure unless the evaluation itself met these conditions. ``Investigator-applied'' means that the research team deliberately changed a network control or configuration during the reported evaluation, rather than observing routine operator activity or proposing an intended deployment. Sustained live control additionally required production-network operation, autonomous or closed-loop action, repeated cycles, service-impact monitoring, and documented recovery or rollback. Observational logs, drive-test measurements, and intended deployment claims did not meet either live-control definition.

We assessed each study using the four checks in Figure~\ref{fig:operator_workflow}: acquisition provenance, evaluation validity, operational exposure, and decision relevance. Numerical outcomes were grouped by theme and task while preserving source metrics, comparators, settings, horizons, and qualifiers such as ``up to.'' Effect sizes were not pooled because datasets, targets, baselines, horizons, and protocols differed. A material inconsistency was defined as a conflict among a study's abstract, main text, tables, figures, or stated evidence-access conditions that changed the interpretation of a principal result or evidence claim. Such studies remained in descriptive and thematic counts when their eligibility and classification were unaffected, but the conflicting numerical values were excluded from representative quantitative conclusions. Unless another denominator is stated, counts use the 72 retained articles; all corpus proportions are descriptive of this corpus and rounded to one decimal place. Evidence-source and algorithm-family counts may overlap, whereas primary themes and highest exposure levels are mutually exclusive.

\section{Empirical tools and data-acquisition taxonomy}
\label{sec:taxonomy}

The 72 studies use a diverse set of empirical data acquisition pipelines. Four groups recur: tools for drive tests and field logging; proprietary operator feeds containing key performance indicators (KPIs), minimization of drive test (MDT) records, or call detail records (CDRs); simulation, emulation, and digital twin platforms, including hardware in the loop (HIL) setups; and measurement frameworks based on crowdsourcing or application data. Section~\ref{sec:quant_profile} reports their distribution across the corpus.

The source category influences evidence fidelity but does not determine it on its own. Drive tests and operator feeds provide real measurements. Simulations and digital twins may be fully synthetic or may combine generated data with measurements. Studies at the application layer may analyze public traces directly or replay them through an emulator. Section~\ref{sec:quant_profile} applies these distinctions to all 72 studies.

\subsection{Drive-test and field-logging tools}

Two instrumentation approaches recur in drive-test data collection. The first is laptop-controlled vendor software interfaced with a test handset or scanner. TEMS Investigation was used with commercial handsets along a 22\,km route for downlink-throughput prediction~\cite{ELDEEB2025103588} and in integrated wireless sensing and positioning experiments~\cite{Zhang2023}. TEMS was additionally used with a Samsung S5 G900I to collect commercial Long-Term Evolution (LTE) drive-test measurements for joint trajectory and signal-power prediction~\cite{10730786}. Such platforms expose detailed radio measurements but require dedicated equipment, active test routes, cleaning, and synchronization before modeling. The second approach is smartphone-native logging, which reduces the control hardware required at the collection point. Nemo Handy, combined with Prometheus exporters, InfluxDB, Docker, and the Measuring Mobile Broadband Networks in Europe (MONROE) Virtual Node, instruments the 5GENESIS facility for end-to-end experimentation~\cite{AUMAYR2022108740}. The public SigCap Android measurements provide the radio-derived labels and geographic coordinates used for Global Positioning System (GPS)-only network-quality classification~\cite{SAHIN2024104002}. In that study, GPS is the final predictive input, while the quality labels are derived from measured reference signal received power (RSRP), reference signal received quality (RSRQ), and received signal strength indicator (RSSI). The public University College Cork (UCC) 4G and 5G traces later reused for mobility- and handoff-aware adaptive streaming were collected through G-NetTrack drive testing; the later study replays those traces in goDASH/goDASHbed and Mininet rather than recollecting them~\cite{HELMY2025104324}. In both approaches, the acquisition tool affects measurement granularity, cost, geographic coverage, and the reproducibility of the resulting model evaluation.

Table~\ref{tab:field_tools} consolidates the named drive-test and field-logging platforms identified in the retained studies. It distinguishes the acquisition instrument from downstream modeling or replay software so that readers can identify practical options for collecting cellular measurements.

\begin{table}[!tbp]
\centering
\caption{Drive-Test and Field-Logging Platforms}
\label{tab:field_tools}
\scriptsize
\renewcommand{\arraystretch}{1.12}
\begin{tabularx}{\columnwidth}{@{}>{\raggedright\arraybackslash}p{2.25cm}>{\raggedright\arraybackslash}X>{\raggedright\arraybackslash}p{1.6cm}@{}}
\toprule
\textbf{Platform} & \textbf{Collection Role} & \textbf{Reported Use} \\
\midrule
TEMS Investigation / Pocket / Discovery & Handset- or scanner-based route logging of radio KPIs, location, and throughput & \cite{ELDEEB2025103588,Zhang2023,10730786} \\
Nemo Handy & Smartphone field logging within the 5GENESIS measurement stack & \cite{AUMAYR2022108740} \\
G-NetTrack Pro & Smartphone 4G/5G route logging; UCC traces later replayed for streaming evaluation & \cite{HELMY2025104324} \\
SigCap & Android radio and GPS measurement collection for coverage-quality labels & \cite{SAHIN2024104002} \\
Keysight Nemo & Timing-advance and serving-cell drive-test measurements & \cite{Eller20225533} \\
PCTEL MXflex & RF scanning and drive-test measurements for three-dimensional network planning & \cite{Eller2022122182} \\
Network Signal Guru + OCR & Rooted-phone screen logging where no measurement API was available & \cite{KALA2022101586} \\
\bottomrule
\end{tabularx}
\end{table}

\subsection{Operator-proprietary KPI and MDT feeds}

Several studies rely on multi-year or large operator-log collections rather than ad hoc drive tests. One study uses a configuration and KPI dataset from the major U.S.\ carrier AT\&T for confounder-aware configuration tuning~\cite{ge2025iridescence}. Another uses more than four years of daily legacy-LTE evolved Node B (eNodeB)-level logs, spanning 412 eNodeBs in a fixed baseline and up to 898 eNodeBs in an evolving dataset, for concept-drift mitigation~\cite{liu2023leaf}. Real minimization-of-drive-test data from approximately 150 cells across two frequency bands underpin differentiable, gradient-based coverage-and-capacity optimization~\cite{10409170}. By contrast, phone measurements captured through Network Signal Guru and OCR are device-side field logs~\cite{KALA2022101586}, and SyntheticNET MDT is generated by a simulator~\cite{RAZA2025108129}; neither was classified as a proprietary operator feed. In the former study, the logs train the capacity model, while the model-guided LTE-LAA cell-selection procedure is executed with the rooted handsets and evaluated through association time and licensed-network capacity. This execution, rather than the logging process alone, supports its physical-experiment and post-action classification.

\subsection{Simulation, edge-layer, and digital-twin platforms}

Where field access is constrained, studies use simulation, emulation, and digital twins with varying degrees of empirical calibration. MATLAB Monte Carlo experiments support idle-mode-capability energy optimization but use no operational trace or hardware power measurement~\cite{11267495}. OMNeT++ produces the 100{,}000 simulated network states used for graph-neural-network service self-healing~\cite{10234541}. A controlled software radio access network (RAN) testbed measures virtual network function (VNF) resource behavior under emulated radio/core conditions for 6G development and operations (DevOps) profiling~\cite{11192422}. The latter is therefore classified as hybrid controlled emulation/testbed evidence, not as a live-network deployment.

\subsection{Crowdsourced and application-layer measurement}

A final category supplements radio/network measurements with contextual or application-layer signals. An automated web crawler gathering football-match schedules supplies contextual event data for sporadic, high-load traffic forecasting in Milan and Rome~\cite{PIMPINELLA2026112147}. In the adaptive-video study, the public UCC 4G/UCC 5G drive-test traces are replayed through goDASH/goDASHbed and Mininet to evaluate mobility- and handoff-aware streaming; their acquisition provenance remains G-NetTrack drive testing, while the downstream evaluation is application-layer emulation~\cite{HELMY2025104324}.

\subsection{Quantitative profile of data-acquisition tools and datasets}
\label{sec:quant_profile}

For the quantitative profile, we examined the evaluation evidence in each of the 72 included studies. The measured/testbed-only category covers field or operator data, public operational releases, and measurements from physical testbeds when no generated component was used in the evaluation. The generated-only category covers synthetic, simulated, or emulated evidence without a measured component. The hybrid category requires an explicit combination of measurements or a physical testbed with simulation, emulation, synthetic augmentation, trace replay, or generated tests of robustness or scalability. Table~\ref{tab:realvssim} reports the distribution.

\begin{table}[!tbp]
\centering
\caption{Measured/Testbed and Generated Evidence in This Corpus ($n=72$)}
\label{tab:realvssim}
\renewcommand{\arraystretch}{1.25}
\begin{tabularx}{\columnwidth}{@{}>{\raggedright\arraybackslash}X>{\centering\arraybackslash}p{1.4cm}>{\centering\arraybackslash}p{1.4cm}@{}}
\toprule
\textbf{Data Source Category} & \textbf{\# Studies} & \textbf{\% in This Corpus} \\
\midrule
Measured/testbed only (field/operator data, public operational release, or physical testbed) & 37 & 51.4\% \\
Generated only (synthetic, simulated, or emulated evidence without a measured component) & 10 & 13.9\% \\
Hybrid (measured/testbed evidence plus a generated or emulated component) & 25 & 34.7\% \\
\bottomrule
\end{tabularx}
\end{table}

In this corpus, 62 of 72 studies (86.1\%) use measurements or a physical testbed, while 35 (48.6\%) include simulation, synthetic generation, emulation, or trace replay; 25 are hybrid. The measured/testbed category includes field or operator data, laboratory testbeds, and legacy public releases. SyntheticNET MDT~\cite{RAZA2025108129} is generated evidence, whereas the measured-layout/calibrated-MDT simulation in~\cite{10409170} is hybrid.

The 62 studies with a measured or physical-testbed component comprise nine physical experiments, 11 offline studies using the 2013--2014 Telecom Italia release, one Telecom Italia study evaluated through simulation, and 41 other measured-data, operator, crowdsourced, or trace-replay studies. One physical experiment also uses the Telecom Italia resource, preserving the corpus-wide reuse count of 13.

Each study was assigned to its highest exposure: 45 (62.5\%) offline analyses, seven (9.7\%) replay or emulation studies, 11 (15.3\%) simulations or synthetic evaluations, and nine (12.5\%) physical experiments. The simulation category includes the Telecom Italia sleep-control evaluation and the synthesized-workload explainability study because both evaluate their procedures under simulated or generated conditions~\cite{Zhang2024,10430406}. The physical group comprises physical-platform, over-the-air, and field procedures~\cite{Gruner2024ipp,10123939,KALA2022101586,Chen2026356,Tsourdinis2024,Bonati2023,11371610,Akgun2024166298,10689363}; seven of these nine studies also executed an evaluated procedure or action on real equipment. Following the rule defined in Section~\ref{sec:methodology}, operator affiliation or access to operational records alone did not raise the exposure level. The classification therefore reflects what occurred during the evaluation rather than the institutional setting in which the study was conducted.

Four complementary validity indicators were also applied corpus-wide using conservative explicit-reporting rules. External validation required evaluation on a distinct site, dataset, or platform (6/72); temporal or spatial separation required a stated non-random holdout (4/72); uncertainty reporting required a quantitative interval, distribution, or uncertainty-aware estimate (3/72); and post-action evaluation required a network KPI measured after an executed action (4/72). Table~\ref{tab:master_table} identifies the studies satisfying each indicator. These counts are reporting floors: a blank means that the criterion was not explicit enough to code, not that the procedure was necessarily absent.

The stated network generation comprises two studies concerning only LTE or 4G, seven spanning LTE or 4G and later generations, 14 targeting 5G or 5G Advanced, 25 concerning 6G or beyond-5G systems, and 24 that are generation-agnostic. Several later-generation studies rely on LTE evidence~\cite{ELDEEB2025103588,liu2023leaf,KALA2022101586,Phyu20235038,Tsourdinis2024}, while many 6G studies use simulation.

Table~\ref{tab:datasetreuse} groups named tools and empirical resources by whether a direct resource link could be identified in the cited studies or relevant resource-provider pages; generic descriptions such as ``commercial network data'' are excluded.

\begin{table*}[!tbp]
\centering
\caption{Named Empirical Resources and Tools Grouped by Availability of Resource Links}
\label{tab:datasetreuse}
\label{tab:public_links}
\footnotesize
\setlength{\tabcolsep}{4pt}
\renewcommand{\arraystretch}{1.16}
\begin{tabularx}{\textwidth}{@{}>{\raggedright\arraybackslash}p{4.7cm}>{\raggedright\arraybackslash}X>{\centering\arraybackslash}p{0.55cm}>{\raggedright\arraybackslash}p{2.4cm}@{}}
\toprule
\textbf{Resource / tool} & \textbf{Access and resource type} & \textbf{$n$} & \textbf{Study references} \\
\midrule
\multicolumn{4}{@{}l}{\textbf{A. Resources with identifiable links}} \\
\href{https://doi.org/10.7910/DVN/EGZHFV}{Telecom Italia Milan} / \href{https://theodi.fbk.eu/openbigdata/}{Trentino} & Public CDR resource & 13 & \cite{9801679,11048623,9810476,10757334,11126163,11526863,Sun2022441,Bettouche2026,Ma20231909,Abdullah202295908,Zhang2024,10376225,11371610} \\
\href{https://github.com/5genesis}{5GENESIS Open Suite} + Nemo Handy & Open research testbed & 1 & \cite{AUMAYR2022108740} \\
\href{https://www.ucc.ie/en/misl/research/datasets/ivid_4g_lte_dataset/}{UCC 4G} / \href{https://github.com/uccmisl/5Gdataset}{UCC 5G} + G-NetTrack & Public drive-test traces & 1 & \cite{HELMY2025104324} \\
\href{https://gitlab.kn.e-technik.tu-dortmund.de/donext-the-dortmund-dataset/dortmund-dataset-public}{DoNext cellular KPI resource} & Public cellular measurements & 1 & \cite{11363362} \\
\href{https://github.com/GCGImdea/NetPredict-Public/CommNetworks2022}{XMLAD / NetPredict} & Public code; Nokia/MONROE records restricted & 1 & \cite{RAMIREZ2023113} \\
\href{https://www.unb.ca/cic/datasets/ids-2017.html}{CICIDS2017} & Public non-cellular proxy benchmark & 1 & \cite{HOSSAIN2025103674} \\
\href{https://openrangym.com}{OpenRAN Gym / PAWR} & Open RF testbed & 1 & \cite{Bonati2023} \\
\href{https://squid.nt.tuwien.ac.at/gitlab/leller/differentiable_throughput_model}{TU Wien Throughput Model} & Public model/code; MDT restricted & 1 & \cite{10409170} \\
\href{https://squid.nt.tuwien.ac.at/gitlab/leller/ieee_access_deep_learning_network_planner}{TU Wien Network Planner} & Public code; drive tests restricted & 1 & \cite{Eller2022122182} \\
\href{https://squid.nt.tuwien.ac.at/gitlab/leller/ieee_access_enodeb_localization}{TU Wien eNodeB Localization} & Public anonymized measurements/code & 1 & \cite{Eller20225533} \\
\href{https://github.com/Code-Dataset/Integrated-Sensing-And-Positioning}{ISAP} & Public measurements and code & 1 & \cite{Zhang2023} \\
\midrule
\multicolumn{4}{@{}l}{\textbf{B. Resources without a confirmed direct link}} \\
TEMS suite & Commercial drive-test instrument & 3 & \cite{ELDEEB2025103588,Zhang2023,10730786} \\
Barcelona 5G traces & Restricted / access unconfirmed & 1 & \cite{11159494} \\
SigCap Android measurements & Public release reported; direct URL unconfirmed & 1 & \cite{SAHIN2024104002} \\
Amarisoft 5G-SA stack & Controlled RF testbed & 1 & \cite{10689363} \\
Network Signal Guru + OCR & Commercial device-side logger & 1 & \cite{KALA2022101586} \\
Milan/Rome KPI + event records & Restricted operator records & 1 & \cite{PIMPINELLA2026112147} \\
Ethio telecom Net-Eco & Restricted monitoring records & 1 & \cite{AHMED2024100814} \\
\bottomrule
\end{tabularx}
\vspace{1mm}
\begin{minipage}{\textwidth}\footnotesize\emph{Notes:} $n$ denotes the number of retained papers using each resource, not the number of freely available papers. A link may lead to a tool, code repository, public measurements, or testbed; it does not establish access to restricted operator records. ``Without a confirmed direct link'' does not imply that the resource is unavailable. The public cellular-access count of 20 studies is determined separately and is not the sum of the rows shown here.\end{minipage}
\end{table*}

The Telecom Italia Milan/Trentino CDR resource has the highest reuse: 13 of 72 studies (18.1\%) overall and 10 of 14 traffic-forecasting studies (71.4\%)~\cite{9801679,11048623,9810476,10757334,11126163,11526863,Sun2022441,Bettouche2026,Ma20231909,Abdullah202295908,10376225,11371610,Zhang2024}. Its 2013--2014 coverage and inconsistent grids, targets, intervals, and splits support within-resource comparison but not a standardized benchmark of current operation.

Identification does not imply accessibility: the Barcelona orchestration traces remain unavailable~\cite{11159494}, while the public XMLAD/NetPredict code excludes the confidential Nokia/MONROE records~\cite{RAMIREZ2023113}. Apart from limited reuse of TEMS, most measurement pipelines rely on one-off devices, facilities, or operator feeds, constraining acquisition replication.

Geographic coverage extends beyond Europe and the United States to China, Malaysia, Indonesia, South Korea, and Ethiopia, but cross-region validation remains uncommon. Broader campaigns and reusable releases should preserve topology, sampling, service, and operator metadata for transfer analysis.

\subsubsection{Public dataset availability}
\label{sec:public_availability}

Public availability required access to at least one empirical cellular dataset or an openly accessible cellular research testbed used by the study. Code-only repositories, non-cellular proxy datasets, commercial tools, proprietary feeds, ``available on request'' statements, restricted data, and unverified claims did not qualify. This rule identified 20 studies with a confirmed public empirical resource, 46 without one, and six that did not specify access. Table~\ref{tab:datasetreuse} distinguishes public cellular resources from commercial tools, restricted records, code-only releases, and non-cellular proxy benchmarks.

Two studies rely primarily on non-cellular proxy benchmarks: MNIST for blockchain-enabled FL~\cite{WAN2022108671} and CICIDS2017 for intrusion detection~\cite{HOSSAIN2025103674}. These resources are not included in the public-cellular-resource count. Their exclusion is retained only as a sensitivity check on the evidence profile, not as an alternative availability count.

\subsection{Key insights and identified gaps}

Empirical tools and testbeds affect fidelity and scalability through trade-offs among measurement realism, geographic and temporal coverage, repeatability, cost, accessibility, and experimental control. Active drive testing provides direct radio measurements but remains costly and geographically narrow: the TEMS throughput study, for example, covers one 22\,km route over one week and does not report an external route or operator test~\cite{ELDEEB2025103588}. Simulation-based studies can scale scenarios and support controlled fault injection, but they introduce a transfer gap. SyntheticNET-generated MDT supports controlled outage detection and compensation, although the reported compensation remains below the ideal normal-case distribution and is not field validated~\cite{RAZA2025108129}. Operator-proprietary KPI feeds can be large and operationally relevant, yet they are rarely standardized or released, constraining reproducibility and cross-study benchmarking. No single acquisition approach therefore maximizes both operational fidelity and scalable, reproducible collection. This concern is consistent with the dataset-standardization gap identified by Lykakis~et al.~\cite{lykakis2025data} and the concentration quantified in Section~\ref{sec:quant_profile}.

\section{Algorithmic techniques and operational constraints}
\label{sec:algorithms}

The reviewed authors employ classical statistical and machine-learning methods, recurrent and state-space sequence architectures, graph-based and attention mechanisms, Transformer variants, reinforcement-learning (RL) controllers, and federated learning (FL), sometimes in combined pipelines. The comparison considers how these families support proactive network management and which operational constraints limit their use.

\subsection{Classical machine learning and statistical models}

Classical ML and statistical approaches appear in 22 retained studies. They are used for throughput and signal prediction, localization, anomaly detection, and configuration analysis, often with lower implementation complexity than the neural alternatives compared in the same study. In a comparison of 13 models using LTE drive-test measurements, the authors identify tuned gradient boosting as their best-performing model, although several secondary values conflict between the authors' discussion and tables~\cite{ELDEEB2025103588}. Localization across different scenarios uses engineered features with AdaBoost but shows substantial error growth in the most difficult rural and indoor settings~\cite{10628101}. Transparent and computationally lighter models remain useful comparators. Their lower complexity, however, does not remove the need for validation at an external site, uncertainty estimates, or measurements of edge computing cost.

\subsection{Recurrent architectures}

Long short-term memory (LSTM) and gated recurrent unit networks appear repeatedly in studies of mobility, traffic, and signal sequences. Hybrid simulation/over-the-air beam prediction uses a 594-parameter LSTM model and reports a 72.31\% prediction rate, 1\% median throughput degradation, and measured central processing unit (CPU) inference of 40\,$\mu$s~\cite{10123939}. In a cloud-native lifecycle prototype, the authors train an LSTM autoencoder on public Telecom Italia traffic and evaluates orchestration on a three-node Kubernetes deployment with an OpenAirInterface (OAI) 5G Core; FlexRIC appears in the radio-access-network collection architecture, while the anomaly proof of concept reports no labeled accuracy or F1-score benchmark~\cite{11371610}. The authors of a label-conditioned LSTM variational autoencoder estimate missing KPIs with most reported mean absolute percentage error (MAPE) values below 5\% and is externally tested on the public DoNext dataset, but its latent uncertainty correlates poorly with actual error~\cite{11363362}. Channel quality indicator (CQI)-prediction experiments report their strongest stated result for an LSTM variant, but nonstandard accuracy definitions, an ambiguous split, and a contradiction between the described four-month tracking data and the final authors' availability statement limit direct comparison~\cite{HASAN2026112126}.

\subsection{Graph-based and attention mechanisms}

Graph neural networks and attention mechanisms are used to model spatial dependencies among cells. The authors evaluate an ensemble graph attention network (GAT) on synthetic graphs and on 23 hours of operator LTE snapshots containing 244{,}694 cell vertices and 976{,}958 neighbor edges. The reported explanations provide model-based attribution, not proof of physical root cause~\cite{10622099}. A hypergraph attention recurrent network (HARN) reports mean absolute error/root mean square error (MAE/RMSE) improvements over the strongest stated comparator of 1.83\%/5.79\% for short message service (SMS), 3.05\%/11.27\% for calls, and 1.36\%/1.65\% for Internet traffic on public Milan data~\cite{10757334}. A controlled 5G standalone (SA) hardware proof of concept combines a graph convolutional network (GCN) policy, REINFORCE, Bayesian-GNN explanations, and symbolic rules, reporting 96.39\% optimal physical resource block (PRB) allocation accuracy but greater computational demand than some deep Q-network (DQN) comparators~\cite{10689363}.

\subsection{Transformer and state-space sequence models}

Transformer and state-space architectures are evaluated for multi-step and long-horizon prediction. Multi-step trajectory experiments on online-transport GPS traces adapted to Jakarta telecom-user distributions report a configuration-specific best average distance error of 0.818\,m, but no RAN measurement or handover outcome is evaluated~\cite{10909538}. Event-driven causal transformers combining wavelets with Granger-based masking reduce average 30\,s mean squared error (MSE) by 17.6\% versus the next-best compared model. Their Web Real-Time Communication (WebRTC) results use Linux trace playback rather than a live in-vehicle link~\cite{liu2025prophet}.

The available computational results do not support a common ranking of latency or readiness for edge deployment. The authors of the compact beam-change LSTM study report microsecond-scale inference on their CPU implementation~\cite{10123939}, whereas the causal Transformer is evaluated on workstation-class graphics hardware~\cite{liu2025prophet}. HiSTM is a hierarchical Mamba state-space model rather than a Transformer; it reports a 29.4\% reduction in mean absolute error with linear sequence-length scaling, but also identifies memory demand for very long sequences as a limitation~\cite{Bettouche2026}. These results reflect different hardware, workloads, and architectures and should not be pooled as if they describe the same deployment setting. Many studies also omit numerical precision, peak memory on the device, energy per inference, and sustained edge throughput. The corpus therefore cannot establish whether sequence models generally meet or fail the latency and resource requirements of 6G.

\subsection{Reinforcement learning and federated architectures}

Reinforcement learning (RL) and federated learning (FL) address closed-loop or distributed optimization in several studies, although much of the evaluation remains simulated or offline. The reviewed RL methods include tabular Q-learning, REINFORCE, actor--critic, and deep RL methods rather than treating every RL implementation as DRL. Soft actor-critic RL combined with a spike-aware LSTM fulfills 96.2\% of RAN demand and 84.7\% of AI demand in one trace-driven Barcelona scenario. The end-to-end O-RAN controller is simulated rather than deployed in a live near-real-time RAN Intelligent Controller (RIC)~\cite{11159494}. SyntheticNET-based single- and multi-agent actor-critic compensation embeds mean RSRP and Jain fairness in its reward, but remains simulation-only and does not restore the ideal normal-case distribution~\cite{RAZA2025108129}. Fed.JST improves average RMSE by 5.95\% and reduces modeled communication overhead to target accuracy by at least 35.29\%; the paper states that its statistical augmented data are not reconstructable, while full model-parameter transmission and property-inference exposure remain acknowledged risks~\cite{11526863}. A blockchain-enabled, differentially private Wasserstein generative adversarial network (DP-WGAN) framework reports simulated convergence and data-utility gains on MNIST. Because MNIST is a non-cellular image benchmark, this result supports method feasibility rather than cellular deployment validity; blockchain provides tamper evidence and traceability, not automatic prevention of malicious updates~\cite{WAN2022108671}.

Across the nine studies identified as using FL, the evaluated training processes are predominantly offline, centrally emulated, server-executed, or architecturally proposed~\cite{10265809,11146731,11526863,Stenhammar202518049,Ferreira20231907,10430406,WAN2022108671,HOSSAIN2025103674,Phyu20235038}. Fed.JST examines partial participation and client heterogeneity but derives communication time from assumed commercial-link rates~\cite{11526863}. Another sustainability study simulates clients sequentially on one RTX 3090~\cite{11146731}, and the connected-vehicle clustering experiment uses an A100 computing cluster~\cite{Stenhammar202518049}. The multi-slice traffic-forecasting study reports lower communication and computation costs through clustered federated training, but also identifies client dropout and heterogeneous processor, storage, and link capabilities~\cite{Phyu20235038}. None of these evaluations reports sustained asynchronous aggregation across operational RAN-edge devices or a dedicated, empirically tested straggler-recovery policy. The available evidence consequently leaves update staleness, dropout recovery, memory pressure, and energy use under heterogeneous physical clients unresolved.

\subsection{Explainability and trustworthiness}

Several retained studies address the performance-explainability trade-off in AI-native management. The authors' Proxy-Lagrangian in-hoc FL approach maintains their attribution-confidence target with 16.67\% lower reported computation time than its post-hoc comparator, but the evidence uses synthesized slice and over-the-top (OTT) workloads that are not independent and identically distributed (non-IID), together with one chosen confidence measure~\cite{10430406}. Transferable loss meta-learning reports more than 25\% improvement only in selected loss-metric mismatch settings and up to 36\% higher operator revenue in a specific 12-slice admission scenario~\cite{10472616}. These studies make the explainability concern identified in the wider reliability literature~\cite{nurcahyanto2025systematic} concrete, but do not establish universal real-world trustworthiness.

The algorithmic families provide complementary forms of proactive management. Classical models offer comparatively transparent baselines for prediction and diagnosis; recurrent, state-space, graph, attention, and Transformer methods model temporal or spatial dependencies; RL supports sequential control; and FL supports distributed learning without centralizing raw data. Their reported benefits remain conditional on the task, dataset, comparator, and evaluation setting. The recurring operational limitations are incomplete external validation, non-comparable runtime evidence, uncertain edge resource and energy costs, communication and participation overhead in distributed learning, limited explanation validation, and limited evidence connecting decisions to measured service outcomes. The corpus therefore supports task-specific selection rather than a general ranking of algorithmic families.

\section{Network optimization domains and applications}
\label{sec:domains}

This section examines how the techniques in Section~\ref{sec:algorithms} are applied across optimization domains. The four application areas below illustrate interventions in mobility, forecasting, resource management, and resilience, while the seven-theme classification provides the broader synthesis.

For the synthesis, we assigned each study to one of the seven primary themes defined in Table~\ref{tab:theme_definitions}, based on its main optimization objective. The reviewed authors could employ more than one algorithmic family.

\begin{table*}[!tbp]
\centering
\caption{Seven Thematic Clusters Used in the Synthesis}
\label{tab:theme_definitions}
\small
\renewcommand{\arraystretch}{1.15}
\begin{tabularx}{\textwidth}{@{}>{\centering\arraybackslash}p{1.5cm}>{\raggedright\arraybackslash}p{3.8cm}>{\raggedright\arraybackslash}X>{\centering\arraybackslash}p{1.6cm}@{}}
\toprule
\textbf{Theme} & \textbf{Name} & \textbf{Operational Scope} & \textbf{Studies} \\
\midrule
A & PHY layer, energy, and spectrum & Propagation, coverage, signal mapping, positioning, beam prediction, spectrum intelligence, and physical-layer energy efficiency & 12 \\
B & Traffic forecasting & Spatial, temporal, and event-aware prediction of cellular traffic, demand, and load & 14 \\
C & Mobility, handover, and localization & Trajectory and location estimation, mobility prediction, handover support, and migration decisions & 12 \\
D & Resource allocation and RRM & Spectrum, power, physical-resource-block, caching, sleep-mode, and service-resource decisions & 9 \\
E & O-RAN, edge, and MLOps orchestration & Programmable RAN control, xApps, slicing, VNF profiling, crowdsourced automation, edge orchestration, model lifecycle support, and concept-drift management & 10 \\
F & Digital twins, self-healing, and faults & Network-twin calibration, anomaly or outage detection, fault prediction, recovery, and self-healing & 7 \\
G & Explainability, FL, and privacy & Explainable decisions, distributed or federated learning, privacy protection, and trust-oriented evaluation & 8 \\
\bottomrule
\end{tabularx}
\end{table*}

\subsection{Handover and mobility management}

Handover and mobility studies focus on proactive handover management, ping-pong avoidance, and trajectory-aware optimization. A feature-engineered AdaBoost localization model transfers without retraining across several commercial-network scenarios, but its median error deteriorates to 1{,}245.71\,m in the hardest reported scenario~\cite{10628101}. Multi-task learning reduces trajectory and RSRP/RSRQ error relative to single-task baselines on one selected 16{,}632-point urban LTE trajectory. The study does not demonstrate broad multi-city transfer~\cite{10730786}. Autoformer-based mobility and handoff prediction reports no stalls in the selected car/MATH-P adaptive-bitrate tests and higher bitrate in some comparisons, but outcomes vary by mobility, handoff, and adaptive bitrate (ABR) condition in the emulated Dynamic Adaptive Streaming over Hypertext Transfer Protocol (DASH) experiments~\cite{HELMY2025104324}.

\subsection{Throughput and load prediction}

Throughput and load prediction studies connect forecasting directly to capacity-planning decisions. The authors of a 13-model LTE drive-test comparison identify tuned Gradient Boosting as best in their Table~1, with $R^2=0.878$, MAE 688.059\,kbps, and RMSE 1{,}051.689\,kbps. These three Gradient Boosting entries are internally consistent and are not among the conflicting secondary values, which are not used here~\cite{ELDEEB2025103588}. Generative-aided, context-aware forecasting reports best-case gains up to 59.1\% MAPE, but this figure is configuration-specific and some baselines retain lower overall MAPE outside event and busy-hour periods~\cite{PIMPINELLA2026112147}. A deployment-aware predictor with a proximal policy optimization (PPO) controller reports total decision latency below 0.5\,ms in synthetic campus and industrial Internet of Things (IoT) simulations, but gives incompatible headline improvement ranges: 10\%/25\% versus 20 to 30\%/40\% for tail latency and service-level agreement (SLA) violations. Because the source reports inconsistent ranges, neither is treated as definitive here~\cite{11554323}.

\subsection{Spectrum and resource management}

Within spectrum and resource management, studies evaluate dynamic spectrum sharing, beam management, and physical-resource-block allocation. In a fully simulated single-cell integrated sensing and communication (ISAC) model, a lightweight feedforward network reduces average energy per bit by 27.3\% and increases throughput by 62.0\% versus fixed allocation, although projected Arm/neural processing unit (NPU) latency is not measured on edge hardware~\cite{11455053}. A tabular Q-learning study evaluates energy and resource management in simulation, but an internal inconsistency in its reported energy-harvesting range prevents its headline numerical values from being used as representative evidence~\cite{MURTHY2026129619}.

\subsection{Self-healing, fault management, and security}

Network resilience brings together self-healing, anomaly detection, and security, but the evidence spans different validation settings. OMNeT++-based digital-twin experiments report a 22.5\% increase in online load-balance degree and lower simulated delay for graph-driven service redeployment, without physical deployment validation~\cite{10234541}. XMLAD reports 100\% recall against its selected KPI/outlier proxy after full cleaning; because the reference is not an independently labeled set of operational faults, that figure should be interpreted within the study's proxy definition, and the real Nokia/MONROE datasets remain confidential~\cite{RAMIREZ2023113}. In security-oriented work, DINO (self-distillation with no labels)-style self-supervised pretraining with federated averaging reports accuracy gains up to 9\% across labeled-data budgets, but the evaluation reuses CICIDS2017 rather than measurements from a 5G vehicle-to-everything (V2X) network. In that setting, FL keeps raw data local but does not by itself establish a complete privacy guarantee~\cite{HOSSAIN2025103674}.

\subsection{Integrating empirical tools, algorithmic techniques, and optimization domains}
\label{sec:synergy}

Sections~\ref{sec:taxonomy} to~\ref{sec:domains} classify the studies by empirical tool, algorithm family, and optimization domain. Figure~\ref{fig:synergy} compares these classifications. Each study belongs to one evidence category but may use several algorithm families. In this corpus, Theme B contains measured/testbed-only evidence in 12 of 14 studies, whereas Themes D, E, F, and G contain larger shares of hybrid or generated-only evidence. RNN and sequence models occur frequently in traffic forecasting, RL mainly in resource allocation and orchestration, and FL mainly in Themes G and B.

\begin{figure*}[!tbp]
\centering
\begin{minipage}[t]{0.43\textwidth}
\centering
\textbf{\small (a) Evidence type within each theme}\\[2pt]
\begin{tikzpicture}[x=0.28cm,y=0.50cm,font=\scriptsize]
\foreach \theme/\m/\h/\g/\y in {A/8/3/1/6,B/12/2/0/5,C/8/4/0/4,D/1/4/4/3,E/5/4/1/2,F/1/4/2/1,G/2/4/2/0}{
  \node[anchor=east,font=\scriptsize\bfseries] at (-0.4,\y+0.16) {\theme};
  \pgfmathsetmacro{\mh}{\m+\h}
  \pgfmathsetmacro{\mhg}{\m+\h+\g}
  \ifnum\m>0 \draw[fill=teal!55,draw=white] (0,\y) rectangle (\m,\y+0.32);
    \node[font=\tiny\bfseries] at (\m/2,\y+0.16) {\m}; \fi
  \ifnum\h>0 \draw[fill=violet!45,draw=white] (\m,\y) rectangle (\mh,\y+0.32);
    \node[font=\tiny\bfseries] at (\m+\h/2,\y+0.16) {\h}; \fi
  \ifnum\g>0 \draw[fill=orange!55,draw=white] (\mh,\y) rectangle (\mhg,\y+0.32);
    \node[font=\tiny\bfseries] at (\mh+\g/2,\y+0.16) {\g}; \fi
}
\node[anchor=west] at (0,-0.55) {\color{teal!65!black}\rule{0.23cm}{0.20cm}\color{black}\hspace{2pt}Measured/testbed only};
\node[anchor=west] at (0,-1.00) {\color{violet!65!black}\rule{0.23cm}{0.20cm}\color{black}\hspace{2pt}Hybrid};
\node[anchor=west] at (0,-1.45) {\color{orange!75!black}\rule{0.23cm}{0.20cm}\color{black}\hspace{2pt}Generated only};
\end{tikzpicture}
\end{minipage}\hfill
\begin{minipage}[t]{0.55\textwidth}
\centering
\textbf{\small (b) Algorithmic family by theme}\\[2pt]
\begin{tikzpicture}[x=0.67cm,y=0.47cm,font=\scriptsize]
\foreach \x/\theme in {0/A,1/B,2/C,3/D,4/E,5/F,6/G}{\node[font=\scriptsize\bfseries] at (\x+0.5,7.28) {\theme};}
\foreach \row/\label/\vals in {
  6/{ML/Stat.}/{10,0,4,2,3,2,1},
  5/{RNN/Seq.}/{2,10,3,1,5,2,3},
  4/{GNN}/{0,5,0,0,0,1,2},
  3/{Transformer}/{1,2,2,2,1,0,0},
  2/{RL}/{0,0,0,3,2,1,1},
  1/{FL}/{0,3,1,0,0,1,4},
  0/{Other}/{1,0,4,3,4,2,2}}
{
  \node[anchor=east] at (-0.18,\row+0.5) {\label};
  \foreach \v [count=\x from 0] in \vals {
    \pgfmathtruncatemacro{\shade}{8+8*\v}
    \draw[fill=blue!\shade,draw=white] (\x,\row) rectangle (\x+1,\row+1);
    \ifnum\v>6 \node[font=\scriptsize\bfseries,text=white] at (\x+0.5,\row+0.5) {\v};
    \else \node[font=\scriptsize\bfseries,text=black] at (\x+0.5,\row+0.5) {\v}; \fi
  }
}
\end{tikzpicture}
\end{minipage}
\caption{Evidence Type and Algorithm Family by Theme in This Corpus. Panel (a) shows mutually exclusive evidence classes; panel (b) allows multiple algorithm families per study. RL includes tabular and deep methods.}
\label{fig:synergy}
\end{figure*}

Large reusable resources such as the Telecom Italia CDR release are repeatedly paired with graph attention, recurrent, and federated architectures. Their spatial and temporal coverage permits controlled comparisons within the same resource family. Large scale does not necessarily imply public access. For example, the Barcelona traces used by~\cite{11159494} support a replay experiment but are not documented sufficiently to be treated as an open dataset. Several studies using operator scale data report only aggregate properties of their inputs. Smaller measurements from a single route or testbed often use engineered features or lighter models suited to the available campaign. The accessible evidence, feasible model complexity, and strength of the deployment claim are therefore related, but the evidence does not establish that a particular tool determines the choice of model.

Simulation and controlled testbeds are also common where direct experiments could disrupt an operational network. OMNeT++ network states~\cite{10234541} and SyntheticNET MDT~\cite{RAZA2025108129} allow fault injection, coordination among multiple agents, and comparison of recovery policies without intervention in a live network. Controlled radio and core emulation in~\cite{11192422} allows repeatable VNF profiling. The PAWR and OpenRAN Gym study~\cite{Bonati2023} lies between software emulation and field deployment because it also demonstrates portability to physical platforms. This evidence is stronger than software simulation alone, but it does not amount to full field validation or resolve transfer to sustained, decentralized operation across multiple vendors.

Data-driven interventions can support resilient, self-optimizing infrastructure when domain-specific prediction or diagnosis leads to a defined network action, is validated under the intended operating conditions, and is accompanied by monitoring, safeguards, recovery, and adaptation. The reviewed studies demonstrate individual parts of this process across mobility, traffic forecasting, resource allocation, O-RAN orchestration, self-healing, and privacy-aware learning, but not sustained multi-vendor operation as a whole. Progress now depends on linking measurement provenance, calibrated evaluation, algorithm selection, executable control, and recovery across domains instead of optimizing a single model or KPI in isolation.

\section{Discussion}
\label{sec:synthesis}

\subsection{Thematic insights}
\label{sec:thematic}

The seven themes show how the role of data-driven methods changes along the path from observation to network action. PHY-layer, forecasting, and mobility studies primarily improve sensing and estimation; resource-allocation and O-RAN studies move toward decision and execution; digital-twin and self-healing studies address assurance and recovery; and explainability, federation, and privacy introduce constraints across the pathway. Traffic forecasting contains 14 studies; PHY-layer and spectrum intelligence and mobility, handover, and localization contain 12 each; O-RAN and edge orchestration contains 10; and resource allocation contains nine. The remaining studies address digital twins and self-healing and explainability, federated learning, and privacy. Across these themes, four issues recur: provenance and benchmark comparability, transfer beyond a single site or platform, computational and coordination cost, and validation of sustained network action.

\begin{figure*}[!t]
\centering
\begingroup
\resizebox{\textwidth}{!}{%
\begin{tikzpicture}[font=\sffamily]
\definecolor{navy}{RGB}{22,53,86}
\definecolor{phyblue}{RGB}{42,111,170}
\definecolor{predictviolet}{RGB}{116,82,157}
\definecolor{mobilitycyan}{RGB}{38,137,167}
\definecolor{decisionorange}{RGB}{220,128,35}
\definecolor{orchteal}{RGB}{0,137,120}
\definecolor{assurered}{RGB}{180,58,62}
\definecolor{trustindigo}{RGB}{70,75,150}
\definecolor{softgray}{RGB}{241,245,248}
\path[use as bounding box] (-.18,-.05) rectangle (16.05,7.08);

\tikzset{
  lane/.style={rounded corners=8pt,draw=navy!20,line width=.8pt,fill=softgray},
  card/.style={rounded corners=8pt,line width=1.15pt,align=center,inner sep=4pt,fill=white,
    minimum width=3.38cm,minimum height=1.18cm},
  badge/.style={circle,minimum size=.47cm,inner sep=0pt,text=white,font=\bfseries\scriptsize},
  count/.style={rounded corners=3pt,inner xsep=4pt,inner ysep=1.5pt,font=\bfseries\tiny,text=white},
  cardtext/.style={anchor=west,font=\bfseries\fontsize{6.4}{7.1}\selectfont,text=navy,align=left},
  govern/.style={densely dotted,line width=1.2pt,draw=trustindigo!72,line cap=round}
}

\shade[left color=trustindigo!17,right color=orchteal!14,rounded corners=9pt]
  (.18,6.17) rectangle (15.86,6.94);
\node[badge,fill=trustindigo] at (.66,6.56) {G};
\node[anchor=west,font=\bfseries\scriptsize,text=navy] at (1.03,6.56)
  {Explainability, federated learning, and privacy};
\node[anchor=east,font=\bfseries\tiny,text=trustindigo!92!black] at (14.82,6.56)
  {CROSS-LAYER GOVERNANCE};
\node[count,fill=trustindigo] at (15.47,6.56) {$n=8$};
\begin{scope}[shift={(9.15,6.56)},scale=.34]
  \draw[trustindigo!92,line width=1.1pt,rounded corners=2pt] (-.43,-.40) rectangle (.43,.34);
  \draw[trustindigo!92,line width=1.1pt] (-.25,.34) arc (180:0:.25 and .35);
  \fill[orchteal] (0,-.04) circle (.085);
  \draw[orchteal,line width=.9pt] (0,-.10)--(0,-.24);
\end{scope}

\foreach \y/\title in {5.28/{RADIO / SENSING LAYER},3.92/{DATA / INTELLIGENCE LAYER},2.56/{RAN CONTROL LAYER},1.20/{MANAGEMENT LAYER}}{
  \node[lane,minimum width=15.68cm,minimum height=1.14cm] at (8.02,\y) {};
  \node[anchor=west,font=\bfseries\scriptsize,text=navy!88] at (.42,\y+.37) {\title};
}

\draw[govern] (15.20,6.17)--(15.20,1.20);
\foreach \y in {5.28,3.92,2.56,1.20}{
  \draw[govern] (15.20,\y)--(14.68,\y);
  \fill[trustindigo!66] (15.20,\y) circle (.045);
}

\node[card,draw=phyblue!82] (a) at (8.38,5.28) {};
\node[badge,fill=phyblue] at (6.92,5.67) {A};
\node[count,fill=phyblue] at (9.82,5.67) {$n=12$};
\begin{scope}[shift={(7.48,5.15)},scale=.31]
  \draw[navy,line width=1.2pt] (-.55,-.68)--(-.55,.58) (-.85,-.68)--(-.55,-.08)--(-.25,-.68);
  \draw[phyblue,line width=1pt] (-.55,.30) arc (-90:90:.28) (-.55,.10) arc (-90:90:.48);
\end{scope}
\node[cardtext] at (7.92,5.15)
  {PHY, energy, and\\ spectrum};

\node[card,draw=predictviolet!82] (b) at (8.38,3.92) {};
\node[badge,fill=predictviolet] at (6.92,4.31) {B};
\node[count,fill=predictviolet] at (9.82,4.31) {$n=14$};
\begin{scope}[shift={(7.50,3.80)},scale=.31]
  \draw[navy,line width=.9pt] (-1.02,-.48)--(-1.02,.48) (-1.02,-.48)--(1.02,-.48);
  \draw[predictviolet,line width=1.35pt] (-.88,-.18)--(-.48,.18)--(-.10,-.04)--(.22,.31);
  \draw[predictviolet,line width=1.35pt,dashed] (.22,.31)--(.56,.05)--(.90,.39);
\end{scope}
\node[cardtext] at (7.92,3.80) {Traffic forecasting};

\node[card,draw=mobilitycyan!82] (c) at (5.82,2.56) {};
\node[badge,fill=mobilitycyan] at (4.36,2.95) {C};
\node[count,fill=mobilitycyan] at (7.26,2.95) {$n=12$};
\begin{scope}[shift={(4.94,2.44)},scale=.30]
  \draw[mobilitycyan!90,line width=1.25pt,dashed] (-1.00,-.26)..controls(-.50,.52)and(.15,-.52)..(.78,.06);
  \fill[navy] (-1.00,-.26) circle (.12);
  \path[fill=mobilitycyan!82,draw=mobilitycyan!95!black,line width=.8pt] (.86,.39) circle (.28);
  \fill[white] (.86,.39) circle (.095);
  \path[fill=mobilitycyan!82] (.67,.20)--(.86,-.21)--(1.05,.20)--cycle;
\end{scope}
\node[cardtext] at (5.35,2.44)
  {Mobility, handover,\\ and localization};

\node[card,draw=decisionorange!85] (d) at (11.18,2.56) {};
\node[badge,fill=decisionorange] at (9.72,2.95) {D};
\node[count,fill=decisionorange] at (12.62,2.95) {$n=9$};
\begin{scope}[shift={(10.27,2.44)},scale=.30]
  \foreach \x in {-1.0,-.5,0,.5}{\foreach \y in {-.45,0,.45}{
    \draw[navy!55,line width=.55pt,fill=softgray] (\x,\y) rectangle ++(.36,.30);
  }}
  \fill[decisionorange!88] (-.5,0) rectangle ++(.36,.30);
  \fill[orchteal!80] (0,.45) rectangle ++(.36,.30);
  \fill[phyblue!80] (.5,-.45) rectangle ++(.36,.30);
\end{scope}
\node[cardtext] at (10.73,2.44)
  {Resource allocation\\ and RRM};

\node[card,draw=orchteal!84] (e) at (5.82,1.20) {};
\node[badge,fill=orchteal] at (4.36,1.59) {E};
\node[count,fill=orchteal] at (7.26,1.59) {$n=10$};
\begin{scope}[shift={(4.92,1.08)},scale=.29]
  \draw[orchteal!95!black,line width=1pt,rounded corners=2pt,fill=orchteal!12]
    (-.45,.08) rectangle (.45,.58);
  \draw[orchteal!90!black,line width=.75pt] (-.24,.40)--(.22,.40) (-.24,.24)--(.10,.24);
  \draw[orchteal!85,line width=1.05pt] (0,.08)--(0,-.16) (-.72,-.16)--(.72,-.16)
    (-.72,-.16)--(-.72,-.46) (.72,-.16)--(.72,-.46);
  \foreach \x in {-.72,.72}{\draw[navy,line width=.85pt,rounded corners=1pt] (\x-.19,-.69) rectangle (\x+.19,-.46);}
\end{scope}
\node[cardtext] at (5.34,1.08)
  {O-RAN, edge, and\\ MLOps orchestration};

\node[card,draw=assurered!84] (f) at (11.18,1.20) {};
\node[badge,fill=assurered] at (9.72,1.59) {F};
\node[count,fill=assurered] at (12.62,1.59) {$n=7$};
\begin{scope}[shift={(10.24,1.08)},scale=.28]
  \draw[navy,line width=1.05pt] (-1.08,-.58)--(-1.08,.45) (-1.34,-.58)--(-1.08,-.06)--(-.82,-.58);
  \draw[phyblue,line width=.9pt] (-1.08,.21) arc (-90:90:.23) (-1.08,.04) arc (-90:90:.40);
  \draw[assurered!92!black,line width=.9pt,rounded corners=2pt,fill=assurered!8]
    (-.25,-.55) rectangle (1.22,.47);
  \draw[assurered!72,line width=.75pt] (-.05,-.30)--(.25,.02)--(.55,-.11)--(.91,.23);
  \draw[navy!60,line width=.75pt] (-.58,-.12)--(-.36,-.12);
  \fill[orchteal] (-.47,-.12) circle (.07);
\end{scope}
\node[cardtext] at (10.72,1.08)
  {Digital twin, faults,\\ and self-healing};

\end{tikzpicture}%
}
\endgroup
\caption{Seven Themes Mapped to Primary Functional Layers. Dotted links denote cross-layer governance.}
\label{fig:theme_bar}
\end{figure*}
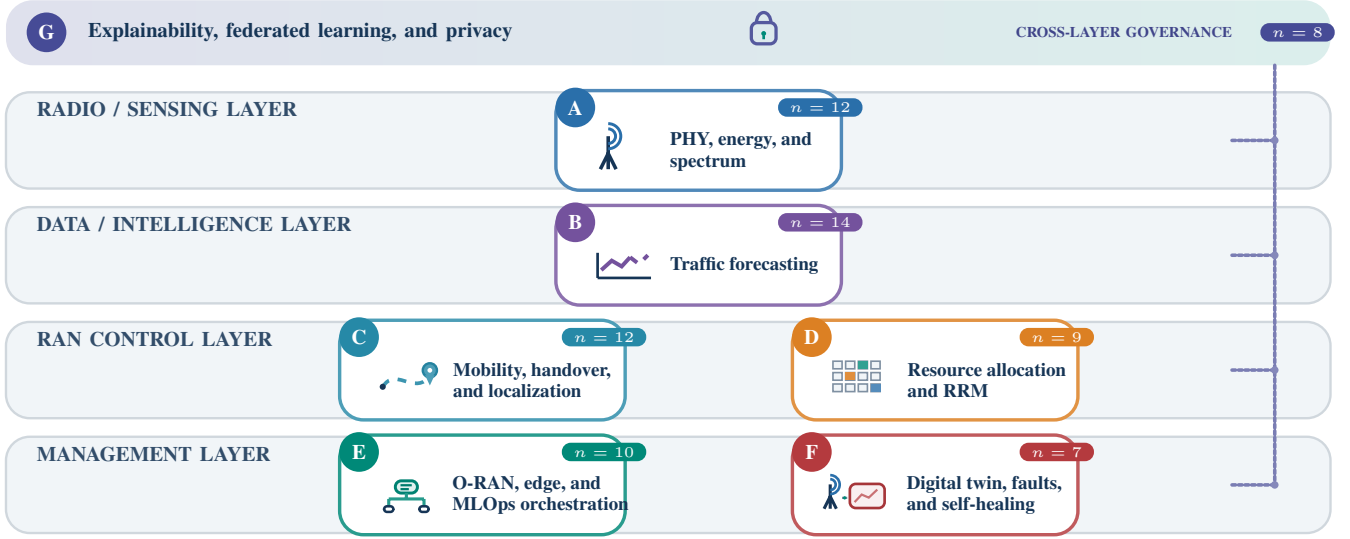

Figure~\ref{fig:theme_bar} positions the themes by their primary functional layer. Theme A is associated with radio sensing; Theme B with data and intelligence; Themes C and D with RAN control; and Themes E and F with management and assurance. Theme G spans the layers because explainability, federation, privacy, and authority can constrain sensing, learning, and control alike.

\subsection{Theme A: PHY-Layer and Spectrum}

The 12 studies largely evaluate propagation, coverage, path loss, signal mapping, throughput or CQI prediction, and beam prediction against measurements or physical experiments~\cite{Liu20224155,Schneider2026,Alimpertis202470,AhmadFauzi202255782,Eller2022122182,Jeske20258022,ELDEEB2025103588,HASAN2026112126,10123939}, with additional work on integrated sensing and positioning~\cite{Zhang2023}, UAV-based private-5G coverage mapping~\cite{Gruner2024ipp}, and energy optimization~\cite{11267495}. Feature efficiency and interpretability recur: one study reduces 64 candidate features to 25 with negligible accuracy loss~\cite{Jeske20258022}, while another uses Explainable Boosting Machines for cross-seasonal analysis~\cite{Schneider2026}. The evidence remains site-, frequency-, and protocol-specific.

\subsection{Theme B: Traffic Forecasting}

Ten of 14 studies use Telecom Italia Milan/Trentino CDR data~\cite{9801679,11048623,9810476,10757334,11126163,11526863,Sun2022441,Bettouche2026,Ma20231909,Abdullah202295908}, with related uses outside the theme~\cite{Zhang2024,10376225,11371610}. The cluster spans graph, hypergraph, recurrent, state-space, federated, multi-task, and weather-aware models. HARN improves prediction error over its stated graph and recurrent baselines~\cite{10757334}, while HiSTM reports a 29.4\% mean-absolute-error reduction and linear sequence-length scaling relative to attention-based alternatives~\cite{Bettouche2026}. A separate study reports a best-case improvement of up to 59.1\% in event-period MAPE by combining operator telemetry, event schedules, and synthetic augmentation; the result is configuration-specific, and some baselines perform better outside event periods~\cite{PIMPINELLA2026112147}. Reuse supports within-resource comparison, but inconsistent grids, targets, horizons, and splits preclude a common ranking and limit evidence of transfer to other network conditions~\cite{JIANG2022117163}.

\subsection{Theme C: Mobility and Localization}

This theme connects trajectory prediction, radio localization, handover support, and mobility-aware service placement, but its metrics are not interchangeable. Multi-step prediction on adapted online-transport GPS traces reports a configuration-specific 0.818\,m average error, yet no RAN measurement or handover outcome is evaluated~\cite{10909538}. A multipath-aware Bayesian probability-hypothesis-density-filter system reports sub-meter localization under its evaluated line-of-sight and non-line-of-sight conditions, while also identifying degradation in severe non-line-of-sight environments and additional edge-processing requirements~\cite{Chen2026356}. A timing-advance estimator identifies up to 95\% of candidate base stations with a mean distance error below 100\,m when confidence-based rejection is allowed~\cite{Eller20225533}. These examples show why headline accuracy must be interpreted together with scenario difficulty, coverage, abstention, and the physical signal used. At the management layer, connected-vehicle clustering reduces MAPE by 8\% and MAE by 7\% in a centrally emulated federated process~\cite{Stenhammar202518049}, trace-driven Lyapunov/particle-swarm-optimization service migration reports energy and latency gains as user load increases~\cite{Zhou2023890}, and drive-test QoE regression supports handover analysis without executing an online decision~\cite{Chmieliauskas202553996}. Cross-city transfer, on-device compute limits, and false-positive or late handover decisions remain insufficiently evaluated across the cluster.

\subsection{Theme D: Resource Allocation}

The methods range from lightweight allocation networks and Q-learning to evolutionary optimization, forecast-driven sleep control, caching, and configuration selection. The reported 27.3\% energy-per-bit reduction~\cite{11455053} is simulation-based; the Q-learning study~\cite{MURTHY2026129619} is retained for categorical mapping, but its conflicting headline values are not used quantitatively. Forecast-driven sleep control~\cite{Zhang2024} and several allocation studies remain simulated or offline, while the physical evaluation in~\cite{KALA2022101586} is bounded to the reported association procedure rather than sustained closed-loop control. The evidence therefore remains strongest for controlled comparison rather than production allocation. A relevant distinction within this theme is whether an optimized quantity is only estimated or whether a configuration is actually applied and its consequences measured. In particular, a simulated reduction in energy use does not by itself establish that activation delays, user association, coverage holes, or traffic variability can be handled during routine operation. These concerns help explain why the physical association experiment and the simulated sleep-control studies support different levels of operational inference, despite addressing related resource-management decisions.

\subsection{Theme E: O-RAN and Orchestration}

OpenRAN Gym demonstrates that containerized data-driven applications developed in emulation can be transferred to heterogeneous physical PAWR platforms~\cite{Bonati2023}. An LTE/OAI testbed controller reports up to fourfold lower latency jitter and nearly fourfold higher throughput than the default equal-allocation baseline~\cite{Tsourdinis2024}. Trace-driven orchestration~\cite{11159494}, controlled VNF emulation~\cite{11192422}, a Kubernetes-based MLOps lifecycle implementation~\cite{11371610}, treatment-conditioned KPI analysis on operational records~\cite{Zhang2026697}, concept-drift management that reduces retraining by up to 76.9\%~\cite{liu2023leaf}, and zero-touch crowdsourcing~\cite{10740263} extend the theme from control to model lifecycle and automation. Cross-vendor generalization and coordination between controllers require further study in the reported settings. Future evaluation should cover interface compatibility, control timing, conflicting actions, safety, and rollback.

\subsection{Theme F: Digital Twins and Self-Healing}

Simulation-only self-healing improves load balance and delay under injected conditions but lacks field validation~\cite{10234541}. An Ethio telecom study uses operational power and environmental measurements, although its abstract and results conflict~\cite{AHMED2024100814}. An indoor over-the-air twin reports up to 14\% higher handover throughput and 50\% lower positioning error at the 80th percentile~\cite{Akgun2024166298}, while federated fraud detection combines proprietary records with simulated federation~\cite{Ferreira20231907}. Together, these studies demonstrate diagnosis, controlled recovery, and physically grounded evaluation, but not sustained autonomous self-healing. Future evaluations should measure calibration drift, fault coverage, service impact, reversibility, and repeated-intervention stability.

\subsection{Theme G: Explainability, FL, and Privacy}

Proxy-Lagrangian FL quantifies an explanation-confidence/computation trade-off on synthesized non-IID workloads~\cite{10430406}. Federated LSTM achieves accuracy comparable to centralized forecasting at lower communication and computation cost, but remains sensitive to dropout and hardware or link heterogeneity~\cite{Phyu20235038}. Explanation quality, formal privacy guarantees, adversarial robustness, and human validation remain uneven; local training alone does not guarantee privacy, and influential inputs do not necessarily reveal physical causes. Future work should test whether explanations remain reliable and aggregation timely enough for network decisions.

Themes A--C emphasize observation or estimation, Themes D and E decisions and execution, Theme F assurance and recovery, and Theme G cross-layer governance. Table~\ref{tab:master_table} records the study-level evidence.

Sensitivity checks did not change the evidence profile. After excluding the 13 Telecom Italia studies ($n=59$), measured-only, hybrid, and generated evidence accounted for 44.1/39.0/16.9\%, while offline, replay, simulation, and physical experiments accounted for 57.6/11.9/16.9/13.6\%. Excluding Theme B ($n=58$) yielded corresponding shares of 43.1/39.7/17.2\% and 55.2/10.3/19.0/15.5\%. Among the 39 studies explicitly addressing 5G, 5G-Advanced, 6G, or beyond-5G, the evidence shares were 38.5/38.5/23.1\% and the exposure shares 46.2/15.4/23.1/15.4\%. Offline analysis remained dominant, and higher exposure remained platform-specific. Five studies contained material internal inconsistencies~\cite{ELDEEB2025103588,11554323,HASAN2026112126,MURTHY2026129619,AHMED2024100814}; conflicting values were excluded from the numerical synthesis but not the categorical counts.

\subsection{Quantifying the application space}
\label{sec:app_space}

Beyond the seven optimization themes in Table~\ref{tab:theme_definitions}, studies were grouped by reported deployment context: vehicular/V2X, industrial or private networks, O-RAN or multi-operator infrastructure, consumer media and OTT, and general macro-cellular optimization. The last category accounts for 50 of 72 studies (69.4\%), while vehicular/V2X and industrial settings are less common. OpenRAN Gym~\cite{Bonati2023} and the LTE/OAI slicing testbed~\cite{Tsourdinis2024} do not demonstrate production-scale multi-vendor operation; in the private-5G campaign~\cite{Gruner2024ipp}, the UAV serves only as a measurement platform.

\onecolumn
\begingroup
\rmfamily\fontsize{9.3}{11.1}\selectfont
\setlength{\tabcolsep}{4.95pt}
\renewcommand{\arraystretch}{1.24}
\setlength{\LTleft}{0pt}\setlength{\LTright}{0pt}
\begin{longtable}{@{}l*{3}{c}cc*{2}{c}c*{7}{c}*{4}{c}@{}}
\caption{Study-Level Evidence, Exposure, Access, Algorithmic, and Validity Classification}\label{tab:master_table}\\
\toprule
\textbf{Study} & \multicolumn{3}{c}{\shortstack{Evidence\\type}} & \shortstack{Evidence\\fidelity} & \shortstack{Operational\\exposure} & \multicolumn{2}{c}{\shortstack{Physical\\evidence}} & \shortstack{Public\\resource} & \multicolumn{7}{c}{\shortstack{Algorithmic\\family}} & \multicolumn{4}{c}{\shortstack{Validation\\indicators}} \\
\cmidrule(lr){2-4}\cmidrule(lr){7-8}\cmidrule(lr){10-16}\cmidrule(l){17-20}
& \textbf{M} & \textbf{G} & \textbf{H} & & & \textbf{Col.} & \textbf{Act.} & & \textbf{CS} & \textbf{SN} & \textbf{GN} & \textbf{TR} & \textbf{RL} & \textbf{FL} & \textbf{O} & \textbf{EV} & \textbf{TS} & \textbf{UQ} & \textbf{PA} \\
\midrule
\endfirsthead
\multicolumn{20}{@{}l}{\textit{Table~\thetable: continued}}\\
\toprule
\textbf{Study} & \multicolumn{3}{c}{\shortstack{Evidence\\type}} & \shortstack{Evidence\\fidelity} & \shortstack{Operational\\exposure} & \multicolumn{2}{c}{\shortstack{Physical\\evidence}} & \shortstack{Public\\resource} & \multicolumn{7}{c}{\shortstack{Algorithmic\\family}} & \multicolumn{4}{c}{\shortstack{Validation\\indicators}} \\
\cmidrule(lr){2-4}\cmidrule(lr){7-8}\cmidrule(lr){10-16}\cmidrule(l){17-20}
& \textbf{M} & \textbf{G} & \textbf{H} & & & \textbf{Col.} & \textbf{Act.} & & \textbf{CS} & \textbf{SN} & \textbf{GN} & \textbf{TR} & \textbf{RL} & \textbf{FL} & \textbf{O} & \textbf{EV} & \textbf{TS} & \textbf{UQ} & \textbf{PA} \\
\midrule
\endhead
\midrule\multicolumn{20}{r}{\footnotesize Continued on next page}\\
\endfoot
\bottomrule
\endlastfoot
\multicolumn{20}{@{}l}{\textbf{Theme A: Physical Layer, Energy Efficiency, and Spectrum Intelligence ($n=12$)}} \\
\studyref{S1}{11267495} &  & \checkmark &  & G & SIM &  &  &  & \checkmark &  &  &  &  &  &  &  &  &  &  \\
\studyref{S2}{Liu20224155} &  &  & \checkmark & C & OFF & \checkmark &  &  & \checkmark &  &  &  &  &  &  &  &  &  &  \\
\studyref{S3}{Schneider2026} & \checkmark &  &  & C & OFF & \checkmark &  &  & \checkmark &  &  &  &  &  &  &  & \checkmark &  &  \\
\studyref{S4}{Alimpertis202470} & \checkmark &  &  & C & OFF &  &  &  & \checkmark &  &  &  &  &  &  &  &  &  &  \\
\studyref{S5}{AhmadFauzi202255782} & \checkmark &  &  & C & OFF & \checkmark &  &  & \checkmark &  &  &  &  &  &  &  &  &  &  \\
\studyref{S6}{Eller2022122182} & \checkmark &  &  & C & OFF & \checkmark &  &  &  &  &  &  &  &  & \checkmark &  & \checkmark &  &  \\
\studyref{S7}{Zhang2023} & \checkmark &  &  & C & OFF & \checkmark &  & \checkmark & \checkmark &  &  &  &  &  &  & \checkmark & \checkmark &  &  \\
\studyref{S8}{Gruner2024ipp} &  &  & \checkmark & P & PHY & \checkmark & \checkmark &  & \checkmark &  &  &  &  &  &  &  &  & \checkmark &  \\
\studyref{S9}{Jeske20258022} & \checkmark &  &  & C & OFF & \checkmark &  &  & \checkmark &  &  &  &  &  &  &  &  &  &  \\
\studyref{S10}{10123939} &  &  & \checkmark & P & PHY & \checkmark & \checkmark &  &  & \checkmark &  & \checkmark &  &  &  &  &  &  &  \\
\studyref{S11}{ELDEEB2025103588} & \checkmark &  &  & C & OFF & \checkmark &  & ? & \checkmark &  &  &  &  &  &  &  &  &  &  \\
\studyref{S12}{HASAN2026112126} & \checkmark &  &  & C & OFF &  &  &  & \checkmark & \checkmark &  &  &  &  &  &  &  &  &  \\
\multicolumn{20}{@{}l}{\textbf{Theme B: Spatiotemporal Traffic Forecasting ($n=14$)}} \\
\studyref{S13}{9801679} & \checkmark &  &  & L & OFF &  &  & \checkmark &  &  & \checkmark &  &  &  &  &  &  &  &  \\
\studyref{S14}{10670094} & \checkmark &  &  & C & OFF &  &  &  &  &  & \checkmark &  &  &  &  & \checkmark &  &  &  \\
\studyref{S15}{11048623} & \checkmark &  &  & L & OFF &  &  & \checkmark &  & \checkmark &  &  &  &  &  &  &  &  &  \\
\studyref{S16}{9810476} & \checkmark &  &  & L & OFF &  &  & \checkmark &  & \checkmark &  &  &  &  &  &  &  &  &  \\
\studyref{S17}{10757334} & \checkmark &  &  & L & OFF &  &  & \checkmark &  & \checkmark & \checkmark &  &  &  &  &  &  &  &  \\
\studyref{S18}{10265809} & \checkmark &  &  & C & OFF &  &  &  &  & \checkmark &  &  &  & \checkmark &  &  &  &  &  \\
\studyref{S19}{11146731} &  &  & \checkmark & C & REP &  &  &  &  &  & \checkmark & \checkmark &  & \checkmark &  &  &  &  &  \\
\studyref{S20}{11126163} & \checkmark &  &  & L & OFF &  &  & \checkmark &  & \checkmark &  &  &  &  &  &  &  &  &  \\
\studyref{S21}{11526863} & \checkmark &  &  & L & OFF &  &  & \checkmark &  &  &  &  &  & \checkmark &  &  &  &  &  \\
\studyref{S22}{PIMPINELLA2026112147} &  &  & \checkmark & C & OFF &  &  &  &  & \checkmark &  &  &  &  &  &  &  &  &  \\
\studyref{S23}{Sun2022441} & \checkmark &  &  & L & OFF &  &  & \checkmark &  & \checkmark &  &  &  &  &  & \checkmark &  &  &  \\
\studyref{S24}{Bettouche2026} & \checkmark &  &  & L & OFF &  &  & \checkmark &  & \checkmark &  &  &  &  &  & \checkmark &  &  &  \\
\studyref{S25}{Ma20231909} & \checkmark &  &  & L & OFF &  &  & \checkmark &  & \checkmark &  & \checkmark &  &  &  &  &  &  &  \\
\studyref{S26}{Abdullah202295908} & \checkmark &  &  & L & OFF &  &  & \checkmark &  & \checkmark & \checkmark &  &  &  &  &  &  &  &  \\
\multicolumn{20}{@{}l}{\textbf{Theme C: Mobility, Handover, and Localization ($n=12$)}} \\
\studyref{S27}{10586988} & \checkmark &  &  & C & OFF & \checkmark &  &  & \checkmark &  &  &  &  &  &  &  &  &  &  \\
\studyref{S28}{10628101} & \checkmark &  &  & C & OFF &  &  &  & \checkmark &  &  &  &  &  &  &  &  &  &  \\
\studyref{S29}{10909538} & \checkmark &  &  & L & OFF &  &  &  &  & \checkmark &  &  &  &  &  &  &  &  &  \\
\studyref{S30}{10730786} & \checkmark &  &  & C & OFF & \checkmark &  &  &  & \checkmark &  &  &  &  &  &  &  &  &  \\
\studyref{S31}{liu2025prophet} &  &  & \checkmark & C & REP &  &  & ? &  &  &  & \checkmark &  &  & \checkmark &  &  &  &  \\
\studyref{S32}{SAHIN2024104002} & \checkmark &  &  & L & OFF &  &  & \checkmark & \checkmark &  &  &  &  &  &  &  &  &  &  \\
\studyref{S33}{HELMY2025104324} &  &  & \checkmark & L & REP &  &  & \checkmark &  & \checkmark &  & \checkmark &  &  &  & \checkmark &  &  &  \\
\studyref{S34}{Zhou2023890} &  &  & \checkmark & C & OFF &  &  &  &  &  &  &  &  &  & \checkmark &  &  &  &  \\
\studyref{S35}{Stenhammar202518049} & \checkmark &  &  & C & REP &  &  &  &  &  &  &  &  & \checkmark &  &  &  &  &  \\
\studyref{S36}{Chen2026356} &  &  & \checkmark & P & PHY & \checkmark &  &  &  &  &  &  &  &  & \checkmark &  &  &  &  \\
\multicolumn{20}{@{}l}{\textbf{Theme C: Mobility, Handover, and Localization (continued)}} \\
\studyref{S37}{Eller20225533} & \checkmark &  &  & C & OFF & \checkmark &  & \checkmark &  &  &  &  &  &  & \checkmark &  &  & \checkmark &  \\
\studyref{S38}{Chmieliauskas202553996} & \checkmark &  &  & C & OFF & \checkmark &  &  & \checkmark &  &  &  &  &  &  &  &  &  &  \\
\multicolumn{20}{@{}l}{\textbf{Theme D: Resource Allocation and Radio Resource Management ($n=9$)}} \\
\studyref{S39}{10409170} &  &  & \checkmark & C & OFF &  &  &  &  &  &  &  &  &  & \checkmark &  &  &  &  \\
\studyref{S40}{11455053} &  & \checkmark &  & G & SIM &  &  &  &  &  &  &  &  &  & \checkmark &  &  &  &  \\
\studyref{S41}{11554323} &  & \checkmark &  & G & SIM &  &  &  &  &  &  &  & \checkmark &  &  &  &  &  &  \\
\studyref{S42}{lekharu2024colabcache} &  &  & \checkmark & C & OFF &  &  &  &  & \checkmark &  &  & \checkmark &  &  &  &  &  &  \\
\studyref{S43}{ge2025iridescence} & \checkmark &  &  & C & OFF &  &  &  & \checkmark &  &  & \checkmark &  &  &  &  &  &  &  \\
\studyref{S44}{KALA2022101586} &  &  & \checkmark & P & PHY & \checkmark & \checkmark &  & \checkmark &  &  &  &  &  &  &  &  &  & \checkmark \\
\studyref{S45}{MURTHY2026129619} &  & \checkmark &  & G & SIM &  &  &  &  &  &  &  & \checkmark &  &  &  &  &  &  \\
\studyref{S46}{Alghazali2025144591} &  & \checkmark &  & G & SIM &  &  &  &  &  &  &  &  &  & \checkmark &  &  &  &  \\
\studyref{S47}{Zhang2024} &  &  & \checkmark & L & SIM &  &  & \checkmark &  &  &  & \checkmark &  &  &  &  &  &  &  \\
\multicolumn{20}{@{}l}{\textbf{Theme E: Open Radio Access Network, Edge, and Machine Learning Operations Orchestration ($n=10$)}} \\
\studyref{S48}{10980086} &  & \checkmark &  & G & SIM &  &  & ? &  & \checkmark &  & \checkmark &  &  &  &  &  &  &  \\
\studyref{S49}{11159494} &  &  & \checkmark & C & REP &  &  &  &  & \checkmark &  &  & \checkmark &  &  &  &  &  &  \\
\studyref{S50}{11192422} &  &  & \checkmark & P & REP &  &  &  &  &  &  &  &  &  & \checkmark &  &  &  &  \\
\studyref{S51}{AUMAYR2022108740} & \checkmark &  &  & P & OFF & \checkmark &  & \checkmark & \checkmark &  &  &  &  &  &  &  &  &  &  \\
\studyref{S52}{Tsourdinis2024} &  &  & \checkmark & P & PHY & \checkmark & \checkmark &  &  &  &  &  &  &  & \checkmark &  &  &  & \checkmark \\
\studyref{S53}{Bonati2023} &  &  & \checkmark & P & PHY & \checkmark & \checkmark & \checkmark &  & \checkmark &  &  & \checkmark &  & \checkmark &  &  &  & \checkmark \\
\studyref{S54}{11371610} & \checkmark &  &  & P & PHY & \checkmark &  & \checkmark &  & \checkmark &  &  &  &  &  &  &  &  &  \\
\studyref{S55}{liu2023leaf} & \checkmark &  &  & C & OFF &  &  &  & \checkmark & \checkmark &  &  &  &  &  &  & \checkmark &  &  \\
\studyref{S56}{10740263} & \checkmark &  &  & C & OFF &  &  &  & \checkmark &  &  &  &  &  &  &  &  &  &  \\
\studyref{S57}{Zhang2026697} & \checkmark &  &  & C & OFF &  &  &  &  &  &  &  &  &  & \checkmark &  &  &  &  \\
\multicolumn{20}{@{}l}{\textbf{Theme F: Digital Twins, Self-Healing, and Fault Management ($n=7$)}} \\
\studyref{S58}{10376225} &  &  & \checkmark & L & OFF &  &  & \checkmark &  & \checkmark &  &  &  &  & \checkmark &  &  &  &  \\
\studyref{S59}{10234541} &  & \checkmark &  & G & SIM &  &  &  &  &  & \checkmark &  &  &  &  &  &  &  &  \\
\studyref{S60}{AHMED2024100814} & \checkmark &  &  & C & OFF &  &  &  &  & \checkmark &  &  &  &  &  &  &  &  &  \\
\studyref{S61}{RAZA2025108129} &  & \checkmark &  & G & SIM &  &  & ? & \checkmark &  &  &  & \checkmark &  &  &  &  &  &  \\
\studyref{S62}{RAMIREZ2023113} &  &  & \checkmark & C & OFF &  &  &  & \checkmark &  &  &  &  &  &  &  &  &  &  \\
\studyref{S63}{Ferreira20231907} &  &  & \checkmark & C & REP &  &  &  &  &  &  &  &  & \checkmark &  &  &  &  &  \\
\studyref{S64}{Akgun2024166298} &  &  & \checkmark & P & PHY & \checkmark & \checkmark &  &  &  &  &  &  &  & \checkmark &  &  &  & \checkmark \\
\multicolumn{20}{@{}l}{\textbf{Theme G: Explainability, Federated Learning, and Privacy ($n=8$)}} \\
\studyref{S65}{10430406} &  & \checkmark &  & G & SIM &  &  &  &  &  &  &  &  & \checkmark &  &  &  &  &  \\
\studyref{S66}{10689363} & \checkmark &  &  & P & PHY & \checkmark & \checkmark &  &  &  & \checkmark &  & \checkmark &  &  &  &  &  &  \\
\studyref{S67}{10622099} &  &  & \checkmark & C & OFF &  &  &  & \checkmark &  & \checkmark &  &  &  &  &  &  &  &  \\
\studyref{S68}{10472616} &  &  & \checkmark & C & OFF &  &  & ? &  & \checkmark &  &  &  &  &  &  &  &  &  \\
\studyref{S69}{11363362} &  &  & \checkmark & C & OFF &  &  & \checkmark &  & \checkmark &  &  &  &  &  & \checkmark &  & \checkmark &  \\
\studyref{S70}{WAN2022108671} &  & \checkmark &  & G & SIM &  &  & ? &  &  &  &  &  & \checkmark & \checkmark &  &  &  &  \\
\studyref{S71}{HOSSAIN2025103674} & \checkmark &  &  & L & OFF &  &  &  &  &  &  &  &  & \checkmark & \checkmark &  &  &  &  \\
\studyref{S72}{Phyu20235038} &  &  & \checkmark & C & OFF &  &  &  &  & \checkmark &  &  &  & \checkmark &  &  &  &  &  \\
\end{longtable}
\par\smallskip{\rmfamily\fontsize{9}{10.8}\selectfont \textbf{Codes.} Evidence: M, measured/testbed only; G, generated only; H, hybrid. Fidelity: G, generated; L, legacy/proxy; C, contemporary real-world; P, physical testbed/over-the-air. Exposure: OFF, offline analysis; REP, replay/emulation; SIM, simulation/synthetic evaluation; PHY, physical experiment. Public resource: \checkmark{} confirmed, ? unconfirmed. Algorithms: CS, classical/statistical; SN, sequential network; GN, graph neural network; TR, Transformer; RL, reinforcement learning; FL, federated learning; O, other. Validity: EV, external validation; TS, temporal/spatial split; UQ, uncertainty quantification; PA, post-action KPI. Blank cells indicate that the attribute was not identified under the stated conservative rule.}
\endgroup

\twocolumn
\FloatBarrier

\section{Challenges, research gaps, and future directions}
\label{sec:gaps}

The synthesis identifies a common transition problem across otherwise different optimization tasks: evidence that supports prediction or controlled feasibility is not automatically evidence of reliable network action. The following gaps concern the conditions needed to cross that boundary, namely reproducible acquisition, transfer from controlled environments, scalable distributed operation, comparable evaluation, and explicit safeguards for deployment.

\subsection{Cross-theme operational gaps}

Across studies with different primary tasks, four gaps recur. The following discussion relates them to the evidence categories, algorithm families, and public-resource classification reported in Section~\ref{sec:quant_profile}, then turns the observations into practical steps for future network-optimization research.

\begin{enumerate}
\item \textbf{Data-curation and instrumentation overhead.} Fifty-two studies (72.2\%) lack a confirmed public dataset or testbed. TEMS requires cleaning and costly drive tests~\cite{ELDEEB2025103588}, XMLAD removes incomplete features~\cite{RAMIREZ2023113}, and NSG measurements require OCR where no API is available~\cite{KALA2022101586}. A reproducible acquisition record should include machine-readable schemas, collection interfaces, synchronization procedures, missing-data rules, and counts of rejected observations. Such records would help researchers separate logging failures from network behavior and reuse measurements without repeating an entire campaign.
\item \textbf{Digital-twin calibration and controlled-evaluation transfer.} Thirty-five studies (48.6\%) include simulation, emulation, generation, or replay. SyntheticNET and OMNeT++ provide controlled evidence without field validation~\cite{RAZA2025108129,10234541}, while the indoor over-the-air twin depends on dedicated accelerator hardware and strict inference latency~\cite{Akgun2024166298}. Transfer evaluation should compare predicted and measured KPIs over time, record telemetry age and calibration error, define recalibration triggers, and test whether decisions remain effective as operating conditions change. These checks would identify when a model can continue, requires recalibration, or should return control to a safe baseline.
\item \textbf{Scalability, non-stationarity, and distributed-system overhead.} Sixteen studies (22.2\%) use FL or RL. Communication is sometimes modeled rather than measured~\cite{11526863}, proxy data are used~\cite{WAN2022108671}, and heterogeneous clients introduce dropout, processing, storage, and link constraints~\cite{Phyu20235038}. Distributed-control experiments need to vary client availability, traffic load, channel quality, compute capacity, drift, and controller conflicts while measuring communication, energy, convergence time, decision latency, dropout, and recovery. Stress tests of this kind would show whether coordination gains persist under realistic network dynamics.
\item \textbf{Absence of standardized, reproducible benchmarks.} Telecom Italia resources appear in 13 studies (18.1\%), yet the selected subsets, aggregation levels, targets, preprocessing, and splits differ. Proprietary feeds and one-off campaigns add further fragmentation. Benchmark suites should provide versioned temporal and spatial splits, common baselines, uncertainty estimates, workload-specific compute measurements, and separate tracks for offline prediction, replay, testbed control, and limited field operation. These tracks would support fair comparison while showing what additional evidence is needed before network action.
\end{enumerate}

\subsubsection{Proposed EFOR reporting checklist}

The recurring evidence gaps provide the basis for EFOR, a six-module checklist covering acquisition provenance, evaluation design, evidence fidelity, computational constraints, distributed coordination, and network-control assurance. Table~\ref{tab:efor_synthesis} applies the relevant indicators across the corpus, with M5 restricted to the nine FL studies. The module-level results show where reporting is already established and where important information is still missing, without collapsing heterogeneous studies into a single readiness score.

\begin{table*}[!t]
\centering
\caption{Proposed EFOR Reporting Checklist and Corpus Indicators}
\label{tab:efor_synthesis}
\scriptsize
\renewcommand{\arraystretch}{1.08}
\setlength{\tabcolsep}{3.2pt}
\arrayrulecolor{blue!60!black}
\begin{tabularx}{\textwidth}{@{}>{\raggedright\arraybackslash}p{2.7cm}>{\raggedright\arraybackslash}p{4.0cm}>{\raggedright\arraybackslash}p{6.0cm}>{\raggedright\arraybackslash}X@{}}
\toprule[1.2pt]
\rowcolor{blue!14}
\textcolor{blue!45!black}{\textbf{Module}} & \textcolor{blue!45!black}{\textbf{Corpus Indicator}} & \textcolor{blue!45!black}{\textbf{Report at Minimum}} & \textcolor{blue!45!black}{\textbf{Operational Purpose}} \\
\midrule
\rowcolor{gray!5} \textbf{M1. Data provenance} & Source, context, period, and scale: 58/72 (80.6\%) & Source; site or operator; period; access; observation unit; exclusions; sample scale & Enables replication and reveals dependence on particular sites or operators. \\
\textbf{M2. Evaluation design} & Explicit temporal or spatial separation: 4/72 (5.6\%) & Split design; leakage controls; missing-data handling; comparator rationale; uncertainty; external validation & Distinguishes robust evaluation from favorable within-sample performance. \\
\rowcolor{gray!5} \textbf{M3. Evidence fidelity} & Calibration or transfer check: 8/35 (22.9\%) & Roles of measured, simulated, replayed, or emulated evidence; calibration method and error & Shows whether controlled evidence transfers to the intended network setting. \\
\textbf{M4. Model and compute} & Workload-specific execution evidence: 12/72 (16.7\%) & Model size and precision; hardware; memory; latency; throughput; measured energy & Tests whether the method can meet its claimed operating constraints. \\
\rowcolor{gray!5} \textbf{M5. Distributed coordination} & Participation or dropout reported: 2/9 FL studies (22.2\%) & Clients; non-IID setting; participation; dropout; aggregation; traffic per round; recovery; privacy & Exposes coordination cost and resilience under heterogeneous participation. \\
\textbf{M6. Deployment and live control} & Post-action network KPI: 4/72 (5.6\%) & Setting; vendor and interface scope; control-loop timing and duration; safeguards; rollback; recovery & Connects an executed decision to service impact and safe recovery. \\
\bottomrule[1.2pt]
\end{tabularx}
\arrayrulecolor{black}
\end{table*}

EFOR complements general ML reporting guidance by adding cellular-network evidence and control requirements. Table~\ref{tab:framework_comparison} shows the distinction.

\begin{table*}[!t]
\centering
\caption{EFOR and Existing ML Reporting Guidance}
\label{tab:framework_comparison}
\scriptsize
\renewcommand{\arraystretch}{1.08}
\setlength{\tabcolsep}{3.2pt}
\arrayrulecolor{blue!60!black}
\begin{tabularx}{\textwidth}{@{}>{\raggedright\arraybackslash}p{3.0cm}>{\raggedright\arraybackslash}p{5.0cm}>{\raggedright\arraybackslash}X@{}}
\toprule[1.2pt]
\rowcolor{blue!14}
\textcolor{blue!45!black}{\textbf{Guidance}} & \textcolor{blue!45!black}{\textbf{Primary Focus}} & \textcolor{blue!45!black}{\textbf{Cellular-Network Detail Added by EFOR}} \\
\midrule
\rowcolor{gray!5} REFORMS~\cite{Kapoor2024REFORMS} & Transparent and reproducible ML-based science & Network-exposure evidence, control authority, rollback, and distributed coordination overhead. \\
Model cards~\cite{Mitchell2019ModelCards} & Intended use, performance conditions, ethical considerations, and caveats & Evidence provenance linked to network actions, operational exposure, recovery, and RAN or orchestration interfaces. \\
\rowcolor{gray!5} NeurIPS reproducibility checklist~\cite{Pineau2021Reproducibility} & Code and data availability, experimental detail, statistics, and reproducibility & Transfer from simulation or replay to testbed or field operation, with coordination cost and rollback. \\
EFOR & Six modules spanning provenance, evaluation, fidelity, compute, coordination, and control & A single reporting path from network evidence to executed action, service impact, safeguards, and recovery. \\
\bottomrule[1.2pt]
\end{tabularx}
\arrayrulecolor{black}
\end{table*}

The controlled 5G-SA radio-resource allocation proof of concept~\cite{10689363} illustrates how the checklist can be used. Its hardware-based experiment provides evidence about fidelity and the execution setting under M3, while the reported resource-block decision addresses the intervention and post-action fields in M6. At the same time, broader operational claims would require service-impact monitoring and recovery procedures. The example therefore preserves the value of the physical demonstration while making the remaining reporting needs visible.

\begin{figure*}[!b]
\centering
\begingroup
\resizebox{\textwidth}{!}{%
\begin{tikzpicture}[font=\sffamily,>=Latex]
\definecolor{navy}{RGB}{20,48,78}
\definecolor{datablue}{RGB}{39,111,171}
\definecolor{evalorange}{RGB}{226,132,39}
\definecolor{fidelityviolet}{RGB}{119,82,160}
\definecolor{computeteal}{RGB}{0,139,126}
\definecolor{distributedblue}{RGB}{36,121,151}
\definecolor{controlred}{RGB}{188,57,64}
\definecolor{softgray}{RGB}{241,245,248}
\path[use as bounding box] (-.18,.55) rectangle (15.90,6.10);

\tikzset{
  card/.style={rounded corners=8pt,line width=1.15pt,minimum width=2.78cm,
    minimum height=1.96cm,align=center,inner sep=4pt,fill=white},
  flow/.style={-{Latex[length=3.4mm,width=2.6mm]},line width=1.45pt,navy!82},
  flowlabel/.style={fill=white,inner xsep=2.7pt,inner ysep=1.2pt,
    font=\tiny\bfseries,text=navy}
}

\shade[left color=datablue!7,right color=fidelityviolet!7,rounded corners=11pt]
  (.10,3.55) rectangle (9.58,5.98);
\node[anchor=west,font=\bfseries\scriptsize,text=navy] at (.38,5.75)
  {EVIDENCE AND VALIDATION};
\shade[left color=computeteal!7,right color=controlred!7,rounded corners=11pt]
  (3.00,.70) rectangle (12.48,3.18);
\node[anchor=east,font=\bfseries\scriptsize,text=navy] at (12.18,2.95)
  {EXECUTION AND ASSURANCE};

\node[card,draw=datablue!80] (m1) at (1.65,4.58) {};
\begin{scope}[shift={(1.65,5.08)},scale=1.12]
  \shade[top color=datablue!26,bottom color=datablue!67,draw=datablue!88,line width=.75pt]
    (-.48,-.30) rectangle (.48,.25);
  \shade[ball color=datablue!38,draw=datablue!88,line width=.75pt] (0,.26) ellipse (.48 and .15);
  \draw[datablue!88,line width=.75pt] (-.48,-.30) arc (180:360:.48 and .15);
\end{scope}
\node[font=\bfseries\scriptsize,text=navy,align=center,text width=2.35cm]
  at ([yshift=-.48cm]m1.center) {M1 Data provenance};

\node[card,draw=evalorange!84] (m2) at (4.72,4.58) {};
\begin{scope}[shift={(4.72,5.08)}]
  \draw[navy!68,line width=.85pt,rounded corners=2pt,fill=evalorange!10]
    (-.52,-.39) rectangle (.52,.40);
  \draw[navy!68,line width=.85pt,rounded corners=1.5pt,fill=white]
    (-.22,.31) rectangle (.22,.49);
  \draw[evalorange!90!black,line width=.9pt] (-.30,.16)--(.29,.16) (-.30,-.03)--(.15,-.03)
    (-.30,-.22)--(.04,-.22);
  \draw[controlred,line width=1.35pt] (.12,-.19)--(.27,-.34)--(.48,-.08);
\end{scope}
\node[font=\bfseries\scriptsize,text=navy,align=center,text width=2.35cm]
  at ([yshift=-.48cm]m2.center) {M2 Evaluation design};

\node[card,draw=fidelityviolet!80] (m3) at (7.79,4.58) {};
\begin{scope}[shift={(7.79,5.02)}]
  \draw[fidelityviolet!88,line width=1.15pt] (-.72,-.25) arc (180:0:.72);
  \draw[navy!60,line width=.8pt] (-.72,-.25)--(.72,-.25);
  \foreach \a in {20,55,90,125,160}{
    \draw[navy!58,line width=.7pt] ({.60*cos(\a)},{-.25+.60*sin(\a)})
      --({.72*cos(\a)},{-.25+.72*sin(\a)});
  }
  \draw[controlred,line width=1.35pt] (0,-.25)--(.34,.22);
  \fill[controlred] (0,-.25) circle (.08);
\end{scope}
\node[font=\bfseries\scriptsize,text=navy,align=center,text width=2.35cm]
  at ([yshift=-.48cm]m3.center) {M3 Evidence fidelity};

\node[card,draw=computeteal!82] (m4) at (4.42,1.68) {};
\begin{scope}[shift={(4.42,2.19)}]
  \draw[computeteal!90!black,line width=.85pt,rounded corners=2pt,fill=computeteal!20]
    (-.50,-.38) rectangle (.50,.38);
  \foreach \x in {-.31,0,.31}{
    \draw[computeteal!90!black,line width=.75pt] (\x,.38)--(\x,.55) (\x,-.38)--(\x,-.55);
  }
  \foreach \y in {-.23,0,.23}{
    \draw[computeteal!90!black,line width=.75pt] (-.50,\y)--(-.67,\y) (.50,\y)--(.67,\y);
  }
  \draw[navy!70,line width=.65pt] (-.25,.16)--(0,0)--(.25,.16) (0,0)--(.25,-.16) (-.25,-.16)--(0,0);
  \fill[navy] (-.25,.16) circle (.055) (-.25,-.16) circle (.055) (0,0) circle (.055)
    (.25,.16) circle (.055) (.25,-.16) circle (.055);
\end{scope}
\node[font=\bfseries\scriptsize,text=navy,align=center,text width=2.35cm]
  at ([yshift=-.48cm]m4.center) {M4 Model and compute};

\node[card,draw=distributedblue!82] (m5) at (7.49,1.68) {};
\begin{scope}[shift={(7.49,2.19)},scale=.92]
  \draw[distributedblue!85,line width=1pt] (-.75,.30)--(0,-.25) (0,.38)--(0,-.25) (.75,.30)--(0,-.25);
  \foreach \x in {-.75,0,.75}{
    \draw[navy!72,line width=.75pt,rounded corners=1pt,fill=distributedblue!12]
      (\x-.18,.16) rectangle (\x+.18,.50);
    \fill[distributedblue] (\x,.22) circle (.035);
  }
  \draw[evalorange!90!black,line width=.85pt,rounded corners=1.5pt,fill=evalorange!22]
    (-.39,-.53) rectangle (.39,-.12);
  \node[font=\bfseries\tiny,text=navy] at (0,-.325) {$\Sigma$};
\end{scope}
\node[font=\bfseries\tiny,text=navy,align=center,text width=2.50cm]
  at ([yshift=-.45cm]m5.center) {M5 Distributed learning\\and coordination};

\node[card,draw=controlred!84] (m6) at (10.56,1.68) {};
\begin{scope}[shift={(10.56,2.19)}]
  \draw[controlred!90!black,line width=.85pt,rounded corners=2pt,fill=controlred!15]
    (-.39,.20) rectangle (.39,.58);
  \draw[controlred!90!black,line width=.75pt] (-.18,.39)--(.18,.39);
  \draw[-{Latex[length=2.1mm,width=1.6mm]},controlred,line width=1.2pt]
    (0,.14)--(0,-.17);
  \draw[navy!78,line width=.9pt,rounded corners=2pt,fill=distributedblue!10]
    (-.61,-.56) rectangle (.61,-.20);
  \draw[navy!62,line width=.7pt] (-.43,-.38)--(.20,-.38);
  \fill[computeteal] (.39,-.38) circle (.055);
\end{scope}
\node[font=\bfseries\scriptsize,text=navy,align=center,text width=2.35cm]
  at ([yshift=-.48cm]m6.center) {M6 Deployment and\\live control};

\draw[flow] (m1.east)--(m2.west) node[midway,above=2pt,flowlabel] {trace};
\draw[flow] (m2.east)--(m3.west) node[midway,above=2pt,flowlabel] {validate};
\draw[flow,rounded corners=8pt] (m3.south)--(7.79,3.34)--(4.42,3.34)--(m4.north);
\node[rounded corners=4pt,fill=navy,text=white,font=\bfseries\tiny,
  inner xsep=5pt,inner ysep=2pt] at (6.18,3.32) {CALIBRATE};
\draw[flow] (m4.east)--(m5.west) node[midway,above=2pt,flowlabel] {execute};
\draw[flow] (m5.east)--(m6.west) node[midway,above=2pt,flowlabel] {coordinate};

\node[rounded corners=9pt,draw=navy,line width=1.2pt,fill=navy,
  minimum width=2.48cm,minimum height=1.84cm,align=center] (claim) at (14.18,1.68) {};
\begin{scope}[shift={(14.18,2.12)}]
  \draw[white!80,line width=.75pt,rounded corners=1.5pt,fill=white]
    (-.53,-.29) rectangle (.53,.31);
  \draw[computeteal,line width=1pt] (-.39,-.11)--(-.16,.05)--(.05,-.04)--(.34,.20);
  \fill[controlred] (.34,.20) circle (.05);
\end{scope}
\node[font=\bfseries\tiny,text=white,align=center,text width=2.02cm]
  at ([yshift=-.44cm]claim.center) {BOUNDED\\DEPLOYMENT CLAIM};
\draw[flow] (m6.east)--(claim.west) node[midway,above=2pt,flowlabel] {control};

\node[rounded corners=5pt,draw=navy!32,fill=white,font=\bfseries\scriptsize,
  text=navy,align=center,inner xsep=8pt,inner ysep=3pt] (rule) at (13.62,4.62)
  {Weakest unsupported link\\bounds the claim};
\draw[-{Latex[length=2.4mm]},dashed,navy!58,line width=.9pt] (rule.south)--(claim.north);

\end{tikzpicture}%
}
\endgroup
\caption{EFOR Reporting Modules and Their Evidence Dependencies.}
\label{fig:bottleneck}
\end{figure*}
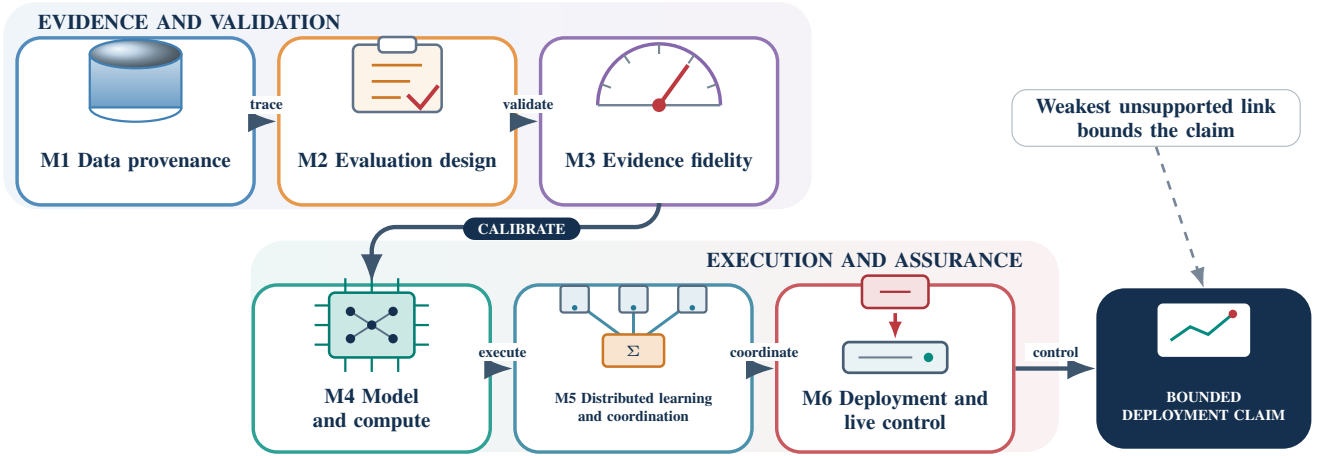

The modules trace a reporting path from provenance and evaluation design through evidence fidelity, execution constraints, coordination, and live control. Figure~\ref{fig:bottleneck} shows their dependencies. The checklist separates controlled feasibility from claims of live effectiveness and makes rollback and recovery explicit.

\subsection{Future research directions}

Future work needs a stronger empirical foundation built on reproducible acquisition records, common temporal and spatial splits, uncertainty reporting, and external validation. Transfer should be tested across times, sites, operators, devices, and vendors, with simulation, replay, and digital twins calibrated against physical operation. Moving in stages from offline replay to hardware-in-the-loop validation, limited field activation, and broader operation would make the evidence gained at each step easier to interpret.

Deployment-oriented studies should evaluate prediction, decision timing, safeguards, drift detection, human intervention, rollback, recovery, and retraining as parts of the same lifecycle. Measurements of latency, memory, sustained throughput, energy, communication per update, participation, dropout, conflicting actions, and recovery should be taken under the workload and hardware claimed for deployment. This would make it possible to judge whether improvements measured in isolation remain useful when the full management process is operating.

\subsection{Limitations}

The empirical-evidence concept block makes the corpus measurement-oriented and may under-represent optimization studies described without measurement-related terminology. Excluding conference, preprint, and industry evidence may also omit operational work reported outside peer-reviewed journals. Screening and classification were conducted by one primary reviewer with co-author supervision, and no formal inter-rater agreement statistic was calculated. Heterogeneous datasets, protocols, targets, and metrics precluded pooled effects. The sensitivity analyses therefore assess the stability of the evidence profile within the reviewed studies. EFOR is derived from this corpus and should be assessed by independent review teams before its consistency is assumed in other network-optimization settings.

\FloatBarrier
\section{Conclusion}
\label{sec:conclusion}

The reviewed literature shows steady progress in prediction, diagnosis, and controlled network optimization, but the strength of each claim depends on how the evidence was obtained and where the method was evaluated. Field and operator measurements preserve operational context but are often narrow or inaccessible. Simulation, replay, emulation, and digital twins provide control and scale, yet their relevance to network operation depends on calibration and transfer across sites, operators, devices, and time.

The algorithm families serve different purposes and cannot be reduced to a single performance hierarchy. Classical methods provide useful baselines, sequence and graph models capture temporal and spatial structure, RL addresses sequential decisions, and FL supports distributed learning. Their practical value still depends on the task, comparator, hardware, implementation cost, and evidence of what happens after a decision is made. More consistent reporting of runtime, energy, communication, external validation, and failure handling would make these results easier to compare and apply.

Reliable network optimization requires a clear connection between validated observations, feasible decisions, executed actions, and measured service outcomes. The EFOR checklist brings these elements together through six reporting modules covering provenance, evaluation, evidence fidelity, compute, coordination, and live control. Used alongside the review's empirical and thematic findings, it provides a practical structure for reporting results and planning evaluations that move beyond predictive accuracy toward dependable network operation.

\section*{Declaration of Generative AI and AI-Assisted Technologies in the Writing Process}
During the preparation of this work, the authors used ChatGPT GPT-5 for language refinement and editing. The authors reviewed and edited the resulting text and accept full responsibility for the content of this publication.

\section*{CRediT Author Statement}
Muhammad Kabeer: Conceptualization, methodology, investigation, formal analysis, visualization, and writing: original draft. Rosdiadee Nordin: Methodology, investigation, formal analysis, validation, and writing: review and editing. Sian Lun Lau: Validation and writing: review and editing.

\section*{Funding}
This research did not receive a specific grant from funding agencies in the public, commercial, or not-for-profit sectors.

\section*{Competing Interests}
The authors declare that they have no known competing financial interests or personal relationships that could have appeared to influence the work reported in this paper.


\end{document}